\documentclass[aps,pra,twocolumn,superscriptaddress]{revtex4-2}

\newcommand{\papertitle}{Phase Sensitivity and Phase Noise of Cantilever-Type Magnetoelastic Sensors Based on the \texorpdfstring{$\Delta$E}{Delta-E} Effect}

\usepackage{amsmath}
\usepackage{amssymb}
\usepackage{sistyle}
\usepackage{hyperref}
\hypersetup{
    unicode=false,
    pdftoolbar=true,
    pdfmenubar=true,
    pdffitwindow=false,
    pdfstartview={FitH},
    pdftitle={\papertitle},
    pdfauthor={Durdaut et al.},
    pdfsubject={\papertitle},
    pdfcreator={Durdaut et al.},
    pdfproducer={Kiel University},
    pdfnewwindow=true,
    colorlinks=true,
    linkcolor=black,
    citecolor=black,
    filecolor=black,
    urlcolor=black
}
\usepackage{graphicx}
\usepackage{subcaption}

\begin{document}


\title{\papertitle}


\author{Phillip Durdaut}
\email[]{pd@tf.uni-kiel.de}
\affiliation{Chair of Microwave Engineering, Institute of Electrical Engineering and Information Technology, Kiel University, Kaiserstr. 2, 24143 Kiel, Germany}

\author{Enrico Rubiola}
\affiliation{FEMTO-ST Institute, Department of Time and Frequency, Universit\'{e} de Bourgogne Franche-Comt\'{e} (UBFC), and CNRS, ENSMM, 26 Rue de l'\'{E}pitaphe, 25000 Besan\c{c}on, France}
\affiliation{Physics Metrology Division, Istituto Nazionale di Ricerca Metrologica (INRiM), Strada Delle Cacce 91, 10135 Torino, Italy}

\author{Jean-Michel Friedt}
\affiliation{FEMTO-ST Institute, Department of Time and Frequency, Universit\'{e} de Bourgogne Franche-Comt\'{e} (UBFC), and CNRS, ENSMM, 26 Rue de l'\'{E}pitaphe, 25000 Besan\c{c}on, France}

\author{Cai M\"uller}
\affiliation{Chair of Nanoscale Magnetic Materials and Magnetic Domains, Institute for Materials Science, Kiel University, Kaiserstr. 2, 24143 Kiel, Germany}

\author{Benjamin Spetzler}
\affiliation{Chair for Multicomponent Materials, Institute for Materials Science, Kiel University, Kaiserstr. 2, 24143 Kiel, Germany}

\author{Christine Kirchhof}
\affiliation{Chair of Inorganic Functional Materials, Institute for Materials Science, Kiel University, Kaiserstr. 2, 24143 Kiel, Germany}

\author{Dirk Meyners}
\affiliation{Chair of Inorganic Functional Materials, Institute for Materials Science, Kiel University, Kaiserstr. 2, 24143 Kiel, Germany}

\author{Eckhard Quandt}
\affiliation{Chair of Inorganic Functional Materials, Institute for Materials Science, Kiel University, Kaiserstr. 2, 24143 Kiel, Germany}

\author{Franz Faupel}
\affiliation{Chair for Multicomponent Materials, Institute for Materials Science, Kiel University, Kaiserstr. 2, 24143 Kiel, Germany}

\author{Jeffrey McCord}
\affiliation{Chair of Nanoscale Magnetic Materials and Magnetic Domains, Institute for Materials Science, Kiel University, Kaiserstr. 2, 24143 Kiel, Germany}

\author{Reinhard Kn\"ochel}
\affiliation{Chair of Microwave Engineering, Institute of Electrical Engineering and Information Technology, Kiel University, Kaiserstr. 2, 24143 Kiel, Germany}

\author{Michael H\"oft}
\affiliation{Chair of Microwave Engineering, Institute of Electrical Engineering and Information Technology, Kiel University, Kaiserstr. 2, 24143 Kiel, Germany}


\date{\today}

\begin{abstract}
Magnetoelastic sensors for the detection of low-frequency and low-amplitude magnetic fields are in the focus of research since more than 30 years. In order to minimize the limit of detection (LOD) of such sensor systems, it is of high importance to understand and to be able to quantify the relevant noise sources. In this contribution, cantilever-type electromechanic and magnetoelastic resonators, respectively, are comprehensively investigated and mathematically described not only with regard to their phase sensitivity but especially to the extent of the sensor-intrinsic phase noise. Both measurements and calculations reveal that the fundamental LOD is limited by additive phase noise due to thermal-mechanical noise of the resonator, i.e. by thermally induced random vibrations of the cantilever, and by thermal-electrical noise of the piezoelectric material. However, due to losses in the magnetic material parametric flicker phase noise arises, limiting the overall performance. In particular it is shown that the LOD is virtually independent of the magnetic sensitivity but is solely determined by the magnetic losses. Instead of the sensitivity, the magnetic losses, represented by the material's effective complex permeability, should be considered as the most important parameter for the further improvement of such sensors in the future. This implication is not only valid for magnetoelastic cantilevers but also applies to any type of magnetoelastic resonator.
\end{abstract}


\maketitle

\section{Introduction}
\label{sec:introduction}
In 1989 Brendel et al. reported on a parasitic influence of magnetic fields on the oscillation frequency of quartz crystal oscillators which could be explained by magnetically induced deformations in the partly ferromagnetic springs used to hold the quartz plate \cite{Bre89,Bre96}. Since then, various micromechanical sensors based on the $\Delta$E effect (Sec.~\ref{subsec:magnetoelastic_sensor_system_delta_e_effect}) have been presented, whose mechanical properties depend on an external magnetic field through interaction with a magnetostrictive layer. Although realizations in the form of highly sensitive magnetoelastic surface acoustic wave delay lines were also presented \cite{Han87,Li12,Elh15,Kit18,Maz18,Wan18a,Sch20}, magnetoelastic sensors are most commonly based on resonant structures \cite{Yos05,Nan13,Kis13,Hui15,Bia16,Li17,Sta17,Ben17,Bia18}, especially cantilevers \cite{Osi96,Goj11,Jah14,Zab15,Par17}, with resonance frequencies in the range between \SI{550}{Hz} and \SI{226}{MHz}.

Besides properties like e.g. dynamic range and frequency bandwidth, the limit of detection (LOD), frequently also referred to as detectivity or equivalent magnetic noise floor, is often considered as one of the most important figures of merit of a magnetic field sensor. Similar to a signal-to-noise ratio (SNR), the LOD is determined by both the sensor's signal, i.e. the sensitivity, as well as by the sensor's noise properties. In the existing articles reporting about magnetoelastic resonators, the focus has mostly been on modified sensor structures and their properties with an emphasis on enhancing the effect, i.e. the detuning of the sensing resonator. Although articles reported on measured values for the limit of detection in the microtesla \cite{Osi96,Yos05,Kis13}, nanotesla \cite{Goj11,Jah14,Hui15,Sta17,Bia18}, and even in the picotesla \cite{Nan13,Zab15,Li17} range, the physical causes for noise in magnetoelastic magnetic field sensors based on the $\Delta$E effect have not been investigated and described yet.

In previous studies, we focused on thermal noise of magnetoelectric cantilevers in passive mode \cite{Dur17a}, on the realization and analysis of low-noise preamplifiers for such sensors \cite{Dur17d}, on noise contributions of the readout electronics \cite{Dur17b,Dur19b} and the suppression of the local oscillator's phase noise in active mode magnetoelastic sensor systems \cite{Dur18}. Based on this, in this paper the influence of the sensor's thermal noise on the phase noise is analyzed both metrologically and analytically. In addition, the impact of losses in the magnetic material on the phase noise characteristics, and thus on the overall sensor performance is shown.

This paper is organized as follows: Sec.~\ref{sec:magnetoelastic_sensor_system} introduces the sensor principle and the actual structure of the magnetoelastic cantilever under investigation. Based on the dynamics of resonant mechanical structures, expressions for the various sensitivities are derived, yielding the overall phase sensitivity. In addition, both an electrical equivalent circuit of the sensor covering for the various loss mechanisms as well a phase detecting readout system is presented. In comparison to previous studies the latter has been modified in order to allow for the neutralization of the sensor's parasitic static capacitance responsible for asymmetric transmission characteristics and a reduced sensitivity. Based on the sensor's loss mechanisms, expressions for thermally induced phase noise are derived and verified by measurements in Sec.~\ref{sec:phase_noise_analysis}. Additional flicker phase noise clearly related to the losses in the magnetic material are traced back to fluctuations of the magnetization. Based on the fluctuation-dissipation theorem analytical expressions for the magnetically induced phase noise as well as for the resulting limit of detection are deduced. This article finishes with a summary of the findings in Sec.~\ref{sec:conclusion}.

\section{Magnetoelastic Sensor System}
\label{sec:magnetoelastic_sensor_system}
\subsection{\texorpdfstring{$\Delta$E}{Delta-E} effect}
\label{subsec:magnetoelastic_sensor_system_delta_e_effect}

The Young's modulus of any material is defined by the ratio between stress $\sigma$ and elastic strain $\varepsilon_{\mathrm{el}}$ that is measured in the direction parallel to the applied stress \cite{Dan09}. However, for magnetic materials the relation ${E_{\mathrm{sat}} = \sigma / \varepsilon_{\mathrm{el}}}$ is only valid for magnetically saturated specimen \cite{Dan09}. In the general case, the problem has to be treated using tensors. As a consequence of the magnetostrictive effect, an additional magnetoelastic strain $\varepsilon_{\mathrm{mel}}$ occurs in magnetic materials \cite[p. 270]{Cul09}. According to ${\Delta\varepsilon_{\mathrm{mel}} = d^{\mathrm{m}} \Delta H}$ (for positive magnetostriction) the magnetoelastic strain directly changes with the magnetic field $H$ and proportionally to the piezomagnetic constant $d^{\mathrm{m}}$ \cite{Cla93} if $\varepsilon_{\mathrm{mel}}$ is linearized around a certain magnetic operating point $H_{\mathrm{bias}}$. Considering both types of elastic strain, the resulting Young's modulus \cite[p. 270]{Cul09} 
\begin{align}
	E(H) = \frac{\sigma}{\varepsilon_{\mathrm{el}} + \varepsilon_{\mathrm{mel}}(H)} \leq E_{\mathrm{sat}}
\end{align}
depends on the magnetic field $H$ and is always lower than the Young's modulus of the same material in magnetic saturation. The magnetically induced change of the Young's modulus in the normalized form
\begin{align}
	\Delta\text{E effect} \equiv \frac{\Delta E}{E} = \frac{E_{\mathrm{sat}} - E}{E} = \frac{\varepsilon_{\mathrm{mel}}(H)}{\varepsilon_{\mathrm{el}}}	
\end{align}
is known as the $\Delta$E effect \cite{Cla93}. In the literature values for ${\Delta E/E}$ as high as approximately \SI{700}{\%} for alloys of terbium-dysprosium (TbDy) \cite{Cla93} and approximately \SI{30}{\%} for an alloy of iron-cobalt-silicon-boron ((Fe$_{90}$Co$_{10}$)$_{78}$Si$_{12}$B$_{10}$) \cite{Lud02b} that is used in this work, respectively, are reported. It should be noted, however, that very large magnetic fields are required to change the Young's modulus in terbium-dysprosium by such a large value. With regard to the use of magnetostrictive materials for sensor applications, it is rather important how strong $E$ is changed by a low amplitude magnetic measurement signal ${B_{\mathrm{x}} = \mu_0 H_{\mathrm{x}}}$ in a certain magnetic operating point ${B_{\mathrm{bias}} = \mu_0 H_{\mathrm{bias}}}$, thus requiring materials with large piezomagnetic constants $d^{\mathrm{m}}$. An overview of piezomagnetic coefficients of various materials can be found in \cite{Don18}. For (Fe$_{90}$Co$_{10}$)$_{78}$Si$_{12}$B$_{10}$ a value of \SI{60}{nm/A} is reported that is only exceeded by (Fe$_{90}$Ga$_{19}$)$_{88}$B$_{12}$ with a value of \SI{151}{nm/A}.

\subsection{Magnetoelastic sensor}
\label{subsec:magnetoelastic_sensors_magnetoelastic_sensor}

\begin{figure}[t]
	\centering
	\includegraphics[width=0.425\textwidth]{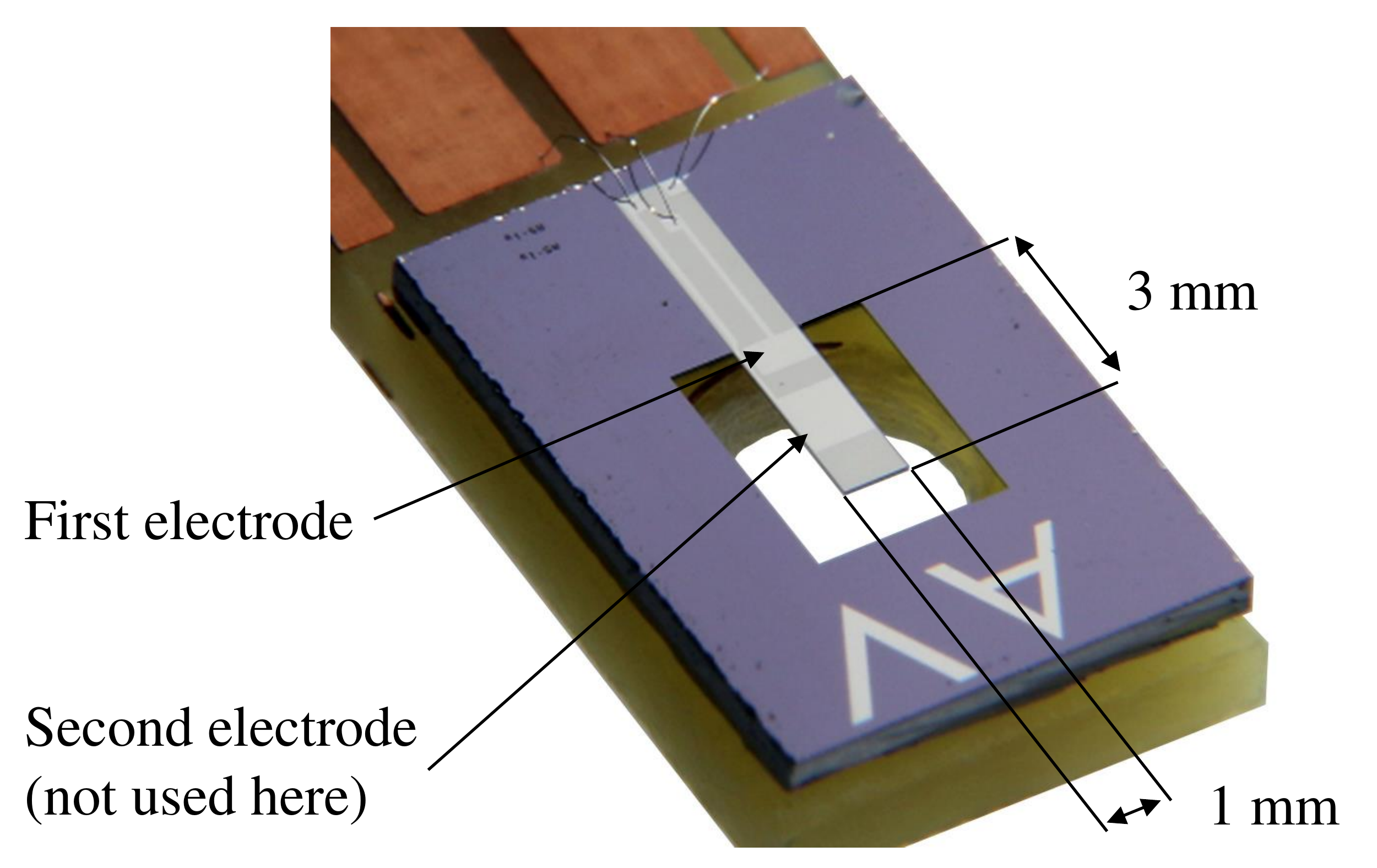}
	\caption{Photograph of the utilized cantilever-type magnetoelastic sensor with a size of \SI{3}{mm} x \SI{1}{mm} mounted to a carrier PCB. The magnetic flux densities $B_{\mathrm{bias}}$ and $B_{\mathrm{x}}$ are applied along the long mechanical axis of the cantilever.}
	\label{fig:sensor}
\end{figure}

The magnetoelastic sensor used for the investigations in this contribution as depicted in Fig.~\ref{fig:sensor} is based on a poly-silicon cantilever of \SI{3}{mm} length, \SI{1}{mm} width and 50~$\mu$m thickness. The lower side is coated with 2~$\mu$m of soft magnetic amorphous metal ((Fe$_{90}$Co$_{10}$)$_{78}$Si$_{12}$B$_{10}$, magnetic easy axis perpendicular to the cantilever's long axis), and 2~$\mu$m of aluminium-nitride (AlN) piezoelectric material \cite{Yar16} are deposited on the cantilever's top. Details about the MEMS fabrication process can be found in \cite{Zab15}. In addition, the sensor offers two independent types of electrodes (see Fig.~\ref{fig:sensor}) that form plate capacitors with the piezoelectric AlN being the dielectric material. The investigations in this contribution focus on the first bending mode for which the first electrode performs best \cite{Zab16}.

\subsection{Resonance detuning and magnetic sensitivity}
\label{subsec:magnetoelastic_sensor_system_resonance_detuning_and_magnetic_sensitivity}

\begin{figure}[t]
	\captionsetup[subfigure]{justification=centering}
	\centering	
	\begin{subfigure}[t]{0.49\textwidth}
		\centering
		\includegraphics[width=1\linewidth]{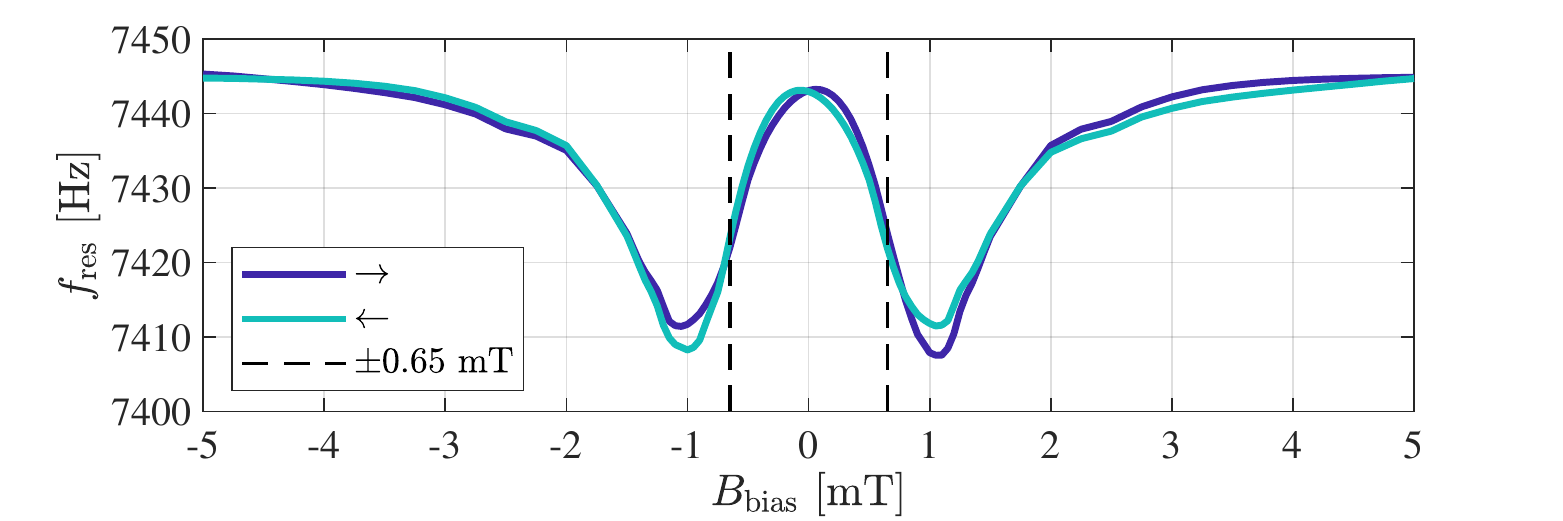}
		\caption{Resonance frequency $f_{\mathrm{res}}$ as a function of the external magnetic bias flux density $B_{\mathrm{bias}}$\\~}
		\label{fig:fres_of_Bbias_for_Vex_10_mV}
	\end{subfigure}
	\begin{subfigure}[t]{0.49\textwidth}
		\centering
		\includegraphics[width=1\linewidth]{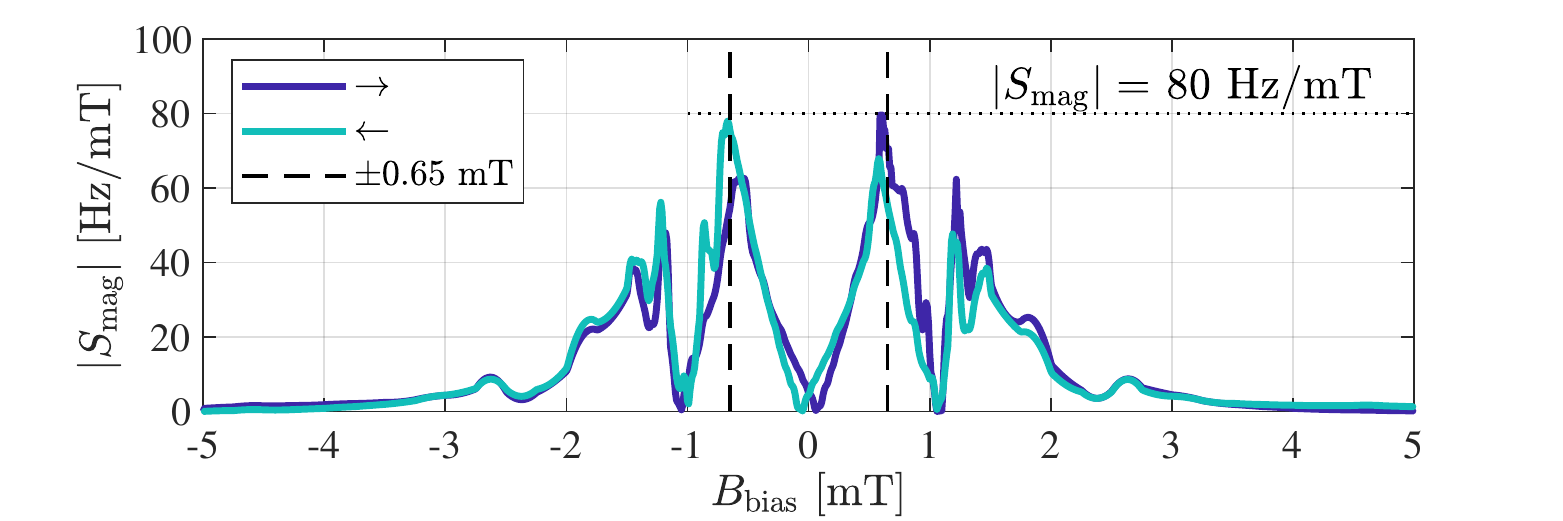}
		\caption{Magnetic sensitivity $S_{\mathrm{mag}}$ as a function of the external magnetic bias flux density $B_{\mathrm{bias}}$\\~}
		\label{fig:Smag_of_Bbias_for_Vex_10_mV}
	\end{subfigure}
	\caption{Detuning of the magnetoelastic cantilever's resonance frequency with an external magnetic flux density (\subref{fig:fres_of_Bbias_for_Vex_10_mV}). High magnetic sensitivities (\subref{fig:Smag_of_Bbias_for_Vex_10_mV}) are reached where the resonance frequency changes particularly strong with an external magnetic flux density, i.e. at ${B_{\mathrm{bias}} = \SI{\pm 0.65}{mT}}$ (vertical dashed lines).}
	\label{fig:fres_and_Smag_of_Bbias_for_Vex_10_mV}
\end{figure}

The resonance frequency of a composite cantilever with $N$ layers is given by \cite{Zab15}
\begin{align}
	f_{\mathrm{res}} = \frac{1}{2 \pi} \frac{\lambda^2}{l^2} \cdot \sqrt{\frac{\sum\limits_{n=1}^{N}~E_n J_n}{\sum\limits_{n=1}^{N}~m_n}},
\end{align}
where $\lambda = 1.875$ is the eigenvalue of the first characteristic bending mode, $l = \SI{3}{mm}$ is the free-standing length of the cantilever, and $E_n$ is the Young's modulus, $J_n$ is the second moment of area, and $m_n$ is the mass per unit length of the $n$-th layer, respectively. Thus, the sensor's resonance frequency is proportional to the square root of the composite's effective Young's modulus $E_{\mathrm{eff}}$
\begin{align}
	f_{\mathrm{res}}(B_{\mathrm{bias}},B_{\mathrm{x}}) \propto \sqrt{E_{\mathrm{eff}}(B_{\mathrm{bias}},B_{\mathrm{x}})}
\end{align}
that depends on an external magnetic field, i.e. on the bias flux density $B_{\mathrm{bias}}$ and on the bias flux density of the measurement signal to be detected $B_{\mathrm{x}}$. As visible from a measurement in Fig.~\ref{fig:fres_of_Bbias_for_Vex_10_mV}, the cantilever's resonance frequency depends on the external magnetic field and changes particularly strong around ${B_{\mathrm{bias}} = \SI{\pm 0.65}{mT}}$. The corresponding slope is known as the magnetic sensitivity towards low amplitude and low frequency magnetic measurement signals $B_{\mathrm{x}}$
\begin{align}
	S_{\mathrm{mag}}(B_{\mathrm{bias}}) = \frac{\partial f_{\mathrm{res}}}{\partial B_{\mathrm{x}}}
	\label{eq:Smag}
\end{align}
for which a maximum value of ${|S_{\mathrm{mag}}| = \SI{80}{Hz/mT}}$ is reached for the sensor under investigation (Fig.~\ref{fig:Smag_of_Bbias_for_Vex_10_mV}). That value is often given normalized to $f_{\mathrm{res}}$ in saturation (here \SI{7445}{Hz}), thus leading to a normalized magnetic sensitivity of ${|S_{\mathrm{mag}}| \approx~1.07~\%f_{\mathrm{res}}/\mathrm{mT}}$ which is a typical value for thin-film magnetoelastic resonators \cite{Yos05,Nan13,Jah14,Zab15,Li17}.

\subsection{Mechanical behavior and electrical sensitivity}
\label{subsec:magnetoelastic_sensor_system_mechanical_behavior_and_electrical_sensitivity}

Generally, the mechanical behavior of a resonant cantilever with the quality factor $Q$ can be described by the unitless frequency response of a simple damped harmonic oscillator \cite[pp. 427]{Hee72} \cite{Kob09}
\begin{align}
	G(f) = \frac{1}{1 - \left(\frac{f}{f_{\mathrm{res}}}\right)^2 + j \frac{f}{f_{\mathrm{res}} Q}} = |G(f)| \cdot \exp\left(j~\gamma(f)\right)
	\label{eq:Gf}
\end{align}
with the magnitude frequency response
\begin{align}
	|G(f)| = \frac{1}{\sqrt{\left(1-\left(\frac{f}{f_{\mathrm{res}}}\right)^2\right)^2 + \left(\frac{f}{f_{\mathrm{res}} Q}\right)^2}}
\end{align}
and the phase response
\begin{align}
	\gamma(f) = \arctan\left( \frac{-\frac{f}{f_{\mathrm{res}}}}{Q \left( 1 -  \left( \frac{f}{f_{\mathrm{res}}} \right)^2 \right)} \right).
	\label{eq:gammaf}
\end{align}
Eq.~\eqref{eq:gammaf} describes the relation between phase and frequency where the slope of $\gamma(f)$ at $f_{\mathrm{res}}$ yields the well-known expression
\begin{align}
	S_{\mathrm{elec}} = \frac{\mathrm{d}\gamma(f)}{\mathrm{d}f}\Bigg|_{f = f_{\mathrm{res}}} = -\frac{2Q}{f_{\mathrm{res}}}
	\label{eq:Selec}
\end{align}
which, in the following, is referred to as the electrical sensitivity $S_{\mathrm{elec}}$ of a resonant sensor in units of ${\mathrm{rad}/\mathrm{Hz}}$ \cite{Dur19a}.

\subsection{Dynamic frequency response and overall phase sensitivity}
\label{subsec:magnetoelastic_sensor_system_dynamic_frequency_response_and_overall_phase_sensitivity}

The higher the quality factor $Q$ of a resonant sensor, the narrower the bandwidth of the characteristic bandpass behavior. Assuming a magnetic measurement signal ${B_{\mathrm{x}}(t) = \hat{B}_{\mathrm{x}} \cos(2 \pi f_{\mathrm{x}} t)}$, the sensor's response to such a signal with the frequency $f_{\mathrm{x}}$ can be determined by replacing $f$ with ${f_{\mathrm{res}} \pm f_{\mathrm{x}}}$ \cite{Kob09} in Eq.~\eqref{eq:Gf}
\begin{align}
	G(f_{\mathrm{res}} \pm f_{\mathrm{x}}) &= \frac{-jQ}{1 \pm \frac{f_{\mathrm{x}}}{f_{\mathrm{res}}} + j\frac{Q f_{\mathrm{x}}^2}{f_{\mathrm{res}}^2} \pm j\frac{2Qf_{\mathrm{x}}}{f_{\mathrm{res}}}}\\
	&\approx \frac{-jQ}{1 \pm j\frac{2 Q f_{\mathrm{x}}}{f_{\mathrm{res}}}}.
\end{align}
Based on that result, an expression for the unitless dynamic sensitivity $S_{\mathrm{dyn}}$ can be deduced
\begin{align}
	S_{\mathrm{dyn}}(f_{\mathrm{x}}) &= \frac{G(f_{\mathrm{res}} \pm f_{\mathrm{x}})}{G(f_{\mathrm{res}})}\\
	&\hspace{-0.5cm}\approx \frac{1}{\sqrt{1 + \left(\frac{f_{\mathrm{x}}}{f_{\mathrm{c}}}\right)^2}} \cdot \exp\left(j~\arctan\left( -\frac{f_{\mathrm{x}}}{f_{\mathrm{c}}} \right)\right)
	\label{eq:Sdynfx}
\end{align}
which exhibits the characteristic of a simple first-order low-pass filter with a cutoff frequency of ${f_{\mathrm{c}} = f_{\mathrm{res}}/(2Q)}$. This result agrees with theoretical expectations in \cite{Mer93} and with measurement results in \cite{Ree16b}. Thus, the overall phase sensitivity of the resonant sensor in units of ${\mathrm{rad}/\mathrm{T}}$ yields
\begin{align}
	S_{\mathrm{PM}}(B_{\mathrm{bias}},f_{\mathrm{x}}) = S_{\mathrm{mag}}(B_{\mathrm{bias}}) \cdot S_{\mathrm{elec}} \cdot S_{\mathrm{dyn}}(f_{\mathrm{x}}).
	\label{eq:SPM}
\end{align}

\subsection{Electrical equivalent circuit and loss mechanisms}
\label{subsec:magnetoelastic_sensor_system_electrical_equivalent_circuit_and_loss_mechanisms}

According to the physical structure of the electromechanical resonator it can be described by an electrical equivalent circuit as depicted in the dashed box in Fig.~\ref{fig:sensor_system} whose element's values can be determined utilizing a conventional impedance analyzer \cite{Zab15,Dur17a,Dur17b}. Based on electromechanical analogies \cite{Blo45} the mechanical structure's resonant behavior is taken into account by an electrical series resonant circuit with the impedance ${Z_{\mathrm{r}} = R_{\mathrm{r}}+R_{\mathrm{mag}} + j \omega L_{\mathrm{r}} + 1/(j \omega C_{\mathrm{r}})}$ where ${\omega = 2 \pi f}$ is the angular frequency. Due to the magnetically induced changes of the resonance frequency ${f_{\mathrm{res}} = 1/(2 \pi \sqrt{L_{\mathrm{r}} C_{\mathrm{r}}})}$, both the inductance $L_{\mathrm{r}}$ as well as the capacitance $C_{\mathrm{r}}$ change with the magnetic field. In parallel to the series resonant circuit the static capacitance due to the electrodes surrounding the piezoelectric material is considered by an additional capacitor with a capacitance of ${C_{\mathrm{ME}} = \SI{44}{pF}}$ for the sensor under investigation.

A piezoelectric cantilever-type magnetoelastic sensor comprises several loss mechanisms that, according to the fluctuation-dissipation theorem, correspond with fluctuations, i.e. with noise. Generally, such losses can be taken into account in an electrical equivalent circuit in the form of dissipative elements, i.e. by resistors.

The predominant loss mechanism of micromechanical cantilevers under atmospheric pressure is air damping, commonly referred to as viscous damping. In addition, e.g. thermoelastic friction intrinsic to the solid structure, support losses, surface losses, and mounting losses (compare \cite{Kir13} for a more detailed analysis for a cantilever like the one investigated here or e.g. \cite{Wei09} for a general overview) may further attenuate the cantilever's deflection, also expressed by its quality factor $Q$. In the electrical equivalent circuit model (dashed box in Fig.~\ref{fig:sensor_system}) these losses are taken into account by the resistance $R_{\mathrm{r}}$. 

\begin{figure}[t]
	\captionsetup[subfigure]{justification=centering}
	\centering	
	\begin{subfigure}[t]{0.49\textwidth}
		\centering
		\includegraphics[width=1.0\linewidth]{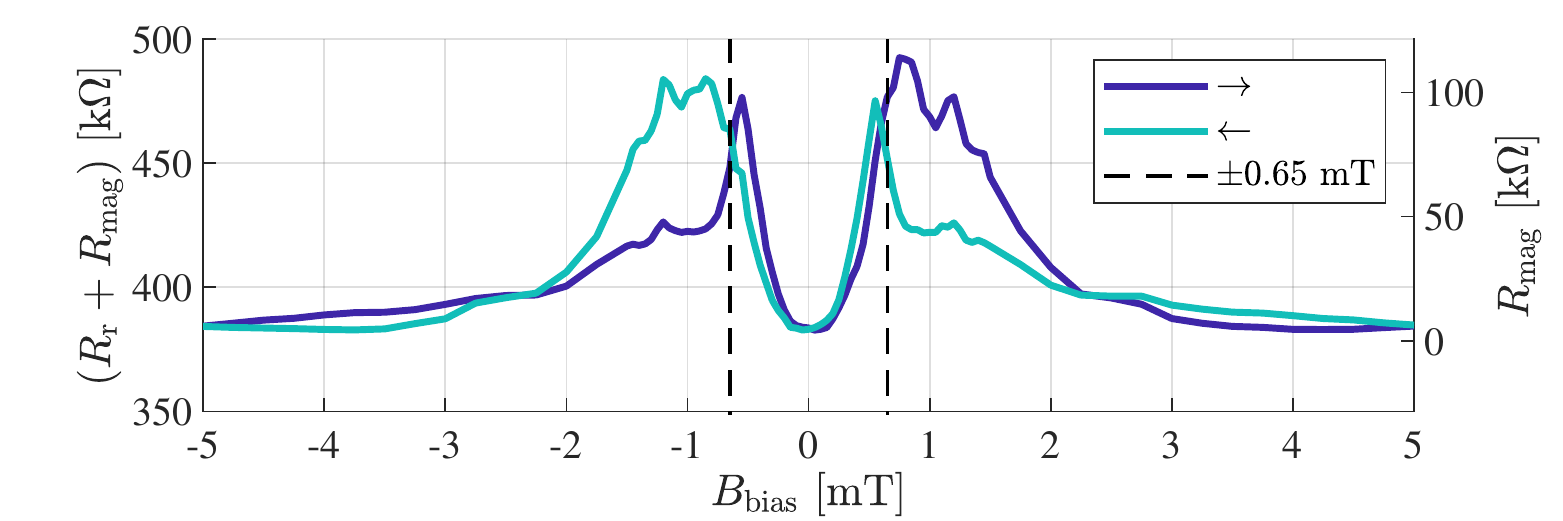}
		\caption{Loss-representing values of the resistances $R_{\mathrm{r}}$ and $R_{\mathrm{mag}}$ as a function of the external magnetic bias flux density $B_{\mathrm{bias}}$\\~}
		\label{fig:RrRmag_of_Bbias}
	\end{subfigure}
	\begin{subfigure}[t]{0.49\textwidth}
		\centering
		\includegraphics[width=1.0\linewidth]{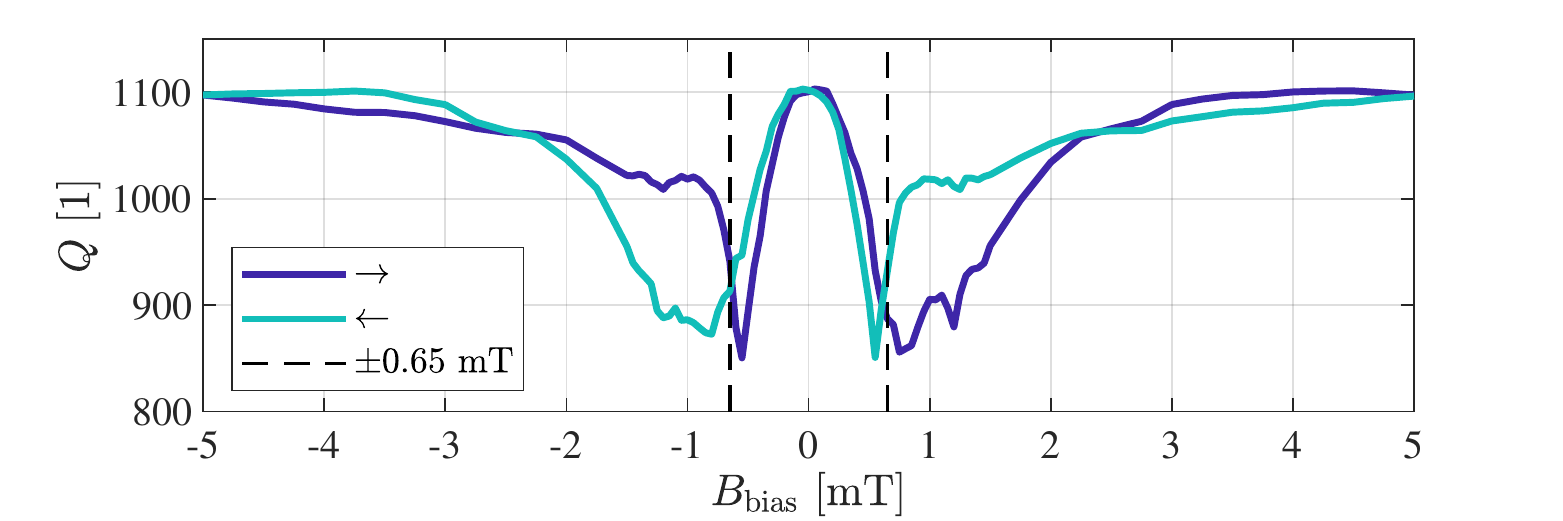}
		\caption{Quality factor $Q$ as a function of the external magnetic bias flux density $B_{\mathrm{bias}}$\\~}
		\label{fig:Q_of_Bbias}
	\end{subfigure}
	\caption{Losses, represented by the resistances $R_{\mathrm{r}}$ and $R_{\mathrm{mag}}$ (\subref{fig:RrRmag_of_Bbias}), and corresponding quality factor $Q$ (\subref{fig:Q_of_Bbias}) as a function of the external magnetic bias flux density $B_{\mathrm{bias}}$ measured for an amplitude of the electrical excitation signal of ${\hat{V}_{\mathrm{ex}} = \SI{100}{mV}}$. The losses are particularly high for bias flux densities that also lead to high magnetic sensitivities $S_{\mathrm{mag}}$ (illustrated by the vertical dashed lines, compare also Fig.~\ref{fig:Smag_of_Bbias_for_Vex_10_mV}) whereas highest values for $Q$ are obtained in magnetic saturation.}
	\label{fig:RrRmag_and_Q_of_Bbias}
\end{figure}

\begin{figure}[t]
	\captionsetup[subfigure]{justification=centering}
	\centering
	\begin{subfigure}[t]{0.49\textwidth}
		\centering
		\includegraphics[width=1.0\linewidth]{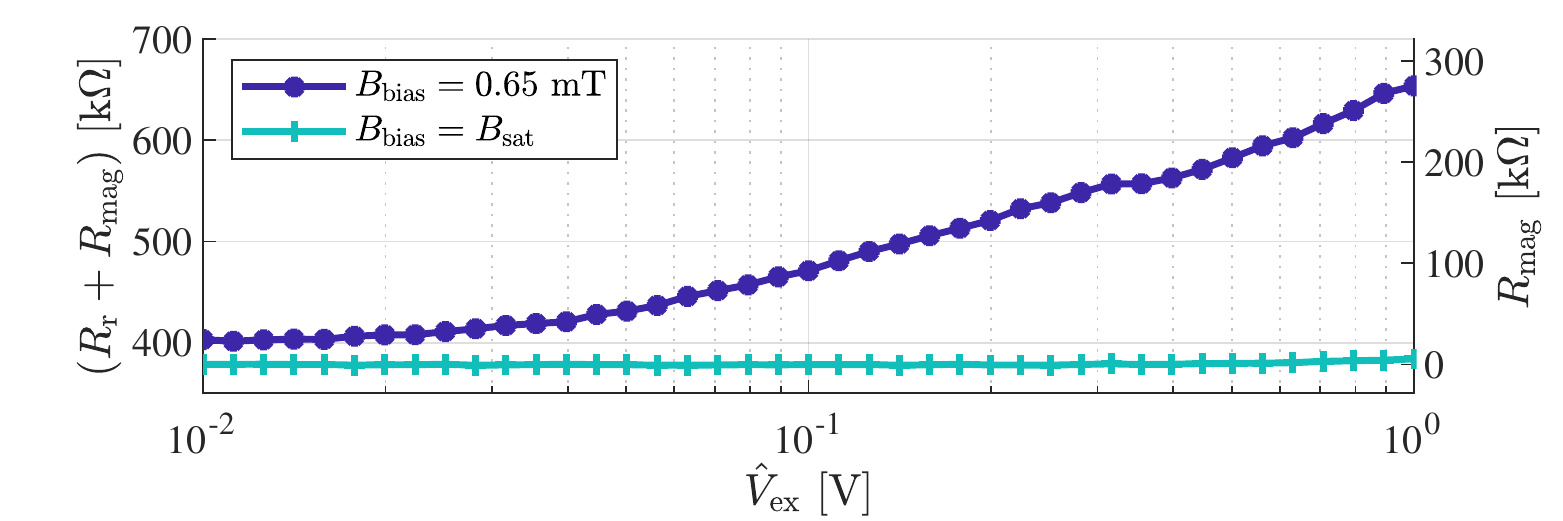}
		\caption{Loss-representing values of the resistances $R_{\mathrm{r}}$ and $R_{\mathrm{mag}}$ as a function of the excitation amplitude $\hat{V}_{\mathrm{ex}}$\\~}
		\label{fig:RrRmag_of_Vexpeak}
	\end{subfigure}
	\begin{subfigure}[t]{0.49\textwidth}
		\centering
		\includegraphics[width=1.0\linewidth]{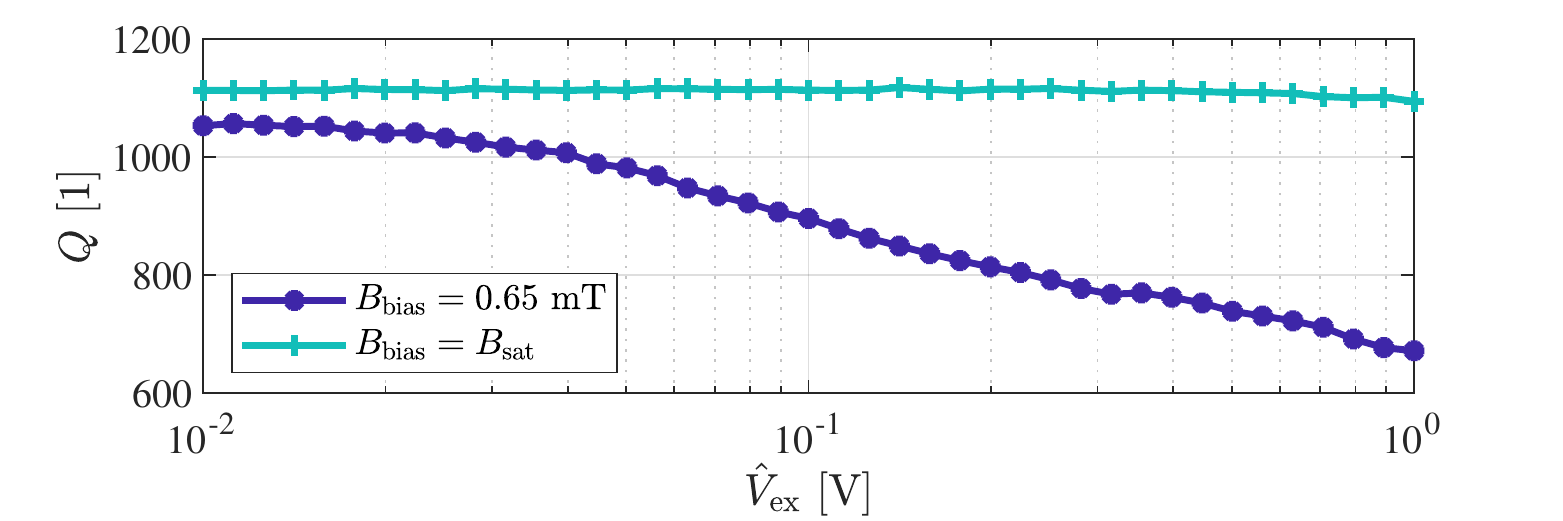}
		\caption{Quality factor $Q$ as a function of the excitation amplitude $\hat{V}_{\mathrm{ex}}$\\~}
		\label{fig:Q_of_Vexpeak}
	\end{subfigure}
	\caption{Losses, represented by the resistances $R_{\mathrm{r}}$ and $R_{\mathrm{mag}}$ (\subref{fig:RrRmag_of_Vexpeak}), and corresponding quality factor $Q$ (\subref{fig:Q_of_Vexpeak}) as a function of the electrical excitation signal's amplitude $\hat{V}_{\mathrm{ex}}$. If the sensor is not magnetically saturated the magnetic losses distinctly increase with the excitation amplitude due to dynamic magnetic hysteresis losses.}
	\label{fig:RrRmag_and_Q_of_Vexpeak}
\end{figure}

For the special case of a magnetoelastic cantilever, additional losses occur as a function of its magnetic state, i.e. as a function of the external magnetic bias flux density $B_{\mathrm{bias}}$ which are considered as an additional resistance $R_{\mathrm{mag}}$. As measurement results in Fig.~\ref{fig:RrRmag_of_Bbias} reveal, these losses are particularly high for bias flux densities that also lead to high magnetic sensitivities $S_{\mathrm{mag}}$ (illustrated by the vertical dashed lines, compare also Fig.~\ref{fig:Smag_of_Bbias_for_Vex_10_mV}). Conversely, this means that the overall quality factor
\begin{align}
	Q(B_{\mathrm{bias}}) = \frac{1}{R_{\mathrm{r}}+R_{\mathrm{mag}}(B_{\mathrm{bias}})} \sqrt{\frac{L_{\mathrm{r}}(B_{\mathrm{bias}})}{C_{\mathrm{r}}(B_{\mathrm{bias}})}}
\end{align}
is also a function of the bias flux density and that $Q$ is lower in the vicinity of the sensor's operating point (here $B_{\mathrm{bias}} = \SI{\pm0.65}{mT}$) than in magnetic saturation (Fig.~\ref{fig:Q_of_Bbias}). Results of a similar series of measurements in Fig.~\ref{fig:RrRmag_and_Q_of_Vexpeak}, but in dependence of the electrical excitation amplitude $V_{\mathrm{ex}}$, clearly confirm the influence of the magnetic state on the losses and on the quality factor, respectively. In fact, the magnetic losses distinctly increase with higher excitation amplitudes in case the magnetic material is not in saturation.

As already hypothesized in \cite{Zab15}, these additional losses can be explained by magnetic hysteresis losses that occur from the periodic bending of the cantilever which, in turn, lead to changes in the magnetization due to the inverse magnetostrictive effect, also referred to as Villari effect \cite{Vil65}. Dynamic magnetic hysteresis losses imply irreversible mechanisms due to domain activity that lead to energy dissipation in the form of heat during each cycle of periodic changes of the magnetization \cite{Ber88,Goo02}. Amongst other loss mechanisms related to the magnetic material like e.g. eddy current losses, the hysteresis losses are considered by the imaginary part ${\mu_{\mathrm{r}}''}$ of the magnetic material's relative permeability ${\mu_{\mathrm{r}} = \mu_{\mathrm{r}}' -j \mu_{\mathrm{r}}''}$ \cite{Goo02}.

Further losses are affiliated to the sensor's plate capacitor, i.e. to the piezoelectric material. These dielectric losses are considered by the loss tangent $\tan \delta_{\mathrm{ME}}$ with reported values for thin-film piezoelectric materials as low as e.g. $2.5 \cdot 10^{-4}$ (aluminium-nitride, AlN) \cite{Yar16a}, $1.3 \cdot 10^{-3}$ (aluminium-scandium-nitride, AlScN) \cite{Fic17}, and $4 \cdot 10^{-3}$ (lead-zirconate-titanate, PZT) \cite{Pio13}. The sensor under investigation exhibits a value of ${\tan \delta_{\mathrm{ME}} = 5 \cdot 10^{-3}}$, thus resulting in a resistance ${R_{\mathrm{ME}} = (\tan \delta_{\mathrm{ME}} \hspace{1mm} \omega C_{\mathrm{ME}})^{-1}}$ in parallel to the static capacitance $C_{\mathrm{ME}}$ and with a value in the vicinity of the resonance frequency of approximately \SI{100}{M}$\Omega$. With a corresponding conductance ${1/R_{\mathrm{ME}}}$ in the nanosiemens range its influence is usually negligible. However, in Sec.~\ref{sec:phase_noise_analysis} it will be shown that the noise associated with these losses might degrade the sensor's performance under certain circumstances.

\subsection{Readout structure}
\label{subsec:magnetoelastic_sensor_system_readout_structure}

\begin{figure*}[t]
	\centering
	\includegraphics[width=1\textwidth]{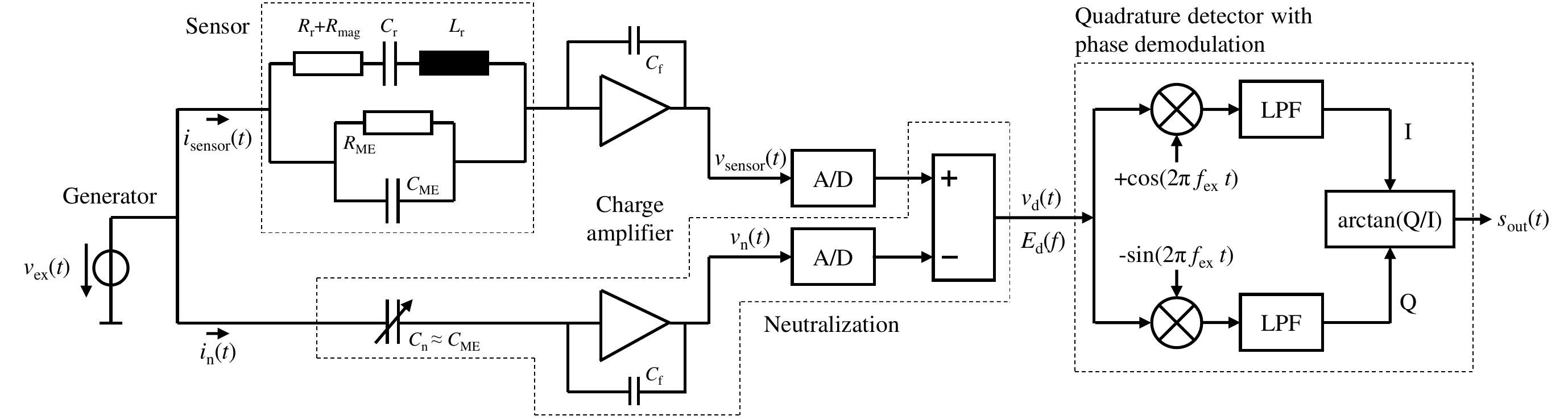}
	\caption{Electrical readout system for the reconstruction of a magnetic measurement signal due to an induced detuning of the resonant sensor. The sensor (upper branch) is driven at its resonance frequency by an electrical excitation signal $v_{\mathrm{ex}}(t)$, leading to a magnetically modulated voltage signal $v_{\mathrm{sensor}}(t)$ at the output of the subsequent charge amplifier. An additional signal branch together with a subtractor is used to neutralize the parasitic effect of the sensor's static capacitance. The phase demodulation is performed in the digital domain by means of a conventional quadrature detector.}
	\label{fig:sensor_system}
\end{figure*}

\begin{figure*}[t]
	\captionsetup[subfigure]{justification=centering}
	\centering
	\begin{subfigure}[t]{0.49\textwidth}
		\centering
		\includegraphics[width=1.0\linewidth]{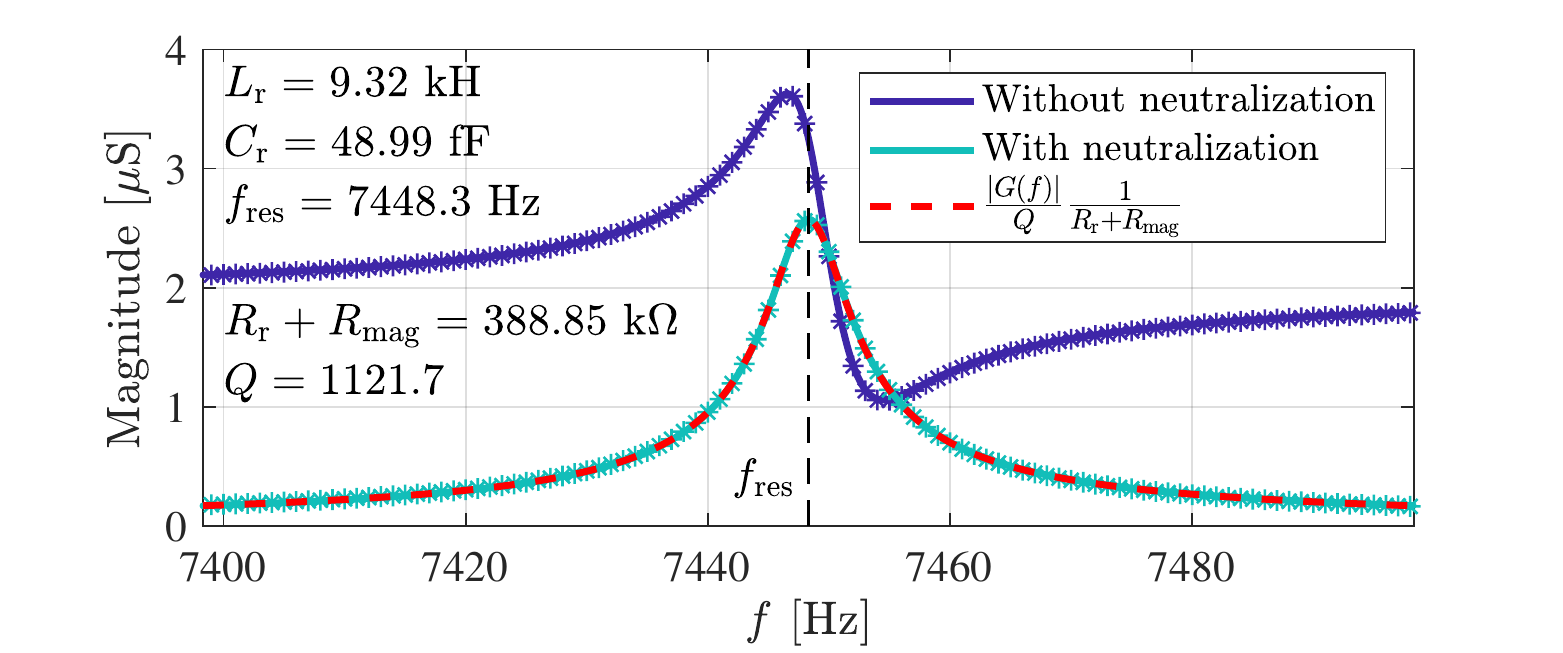}
		\caption{Magnitude of the sensor's admittance\\~}
		\label{fig:admittances_magnitude_of_f}
	\end{subfigure}
	\begin{subfigure}[t]{0.49\textwidth}
		\centering
		\includegraphics[width=1.0\linewidth]{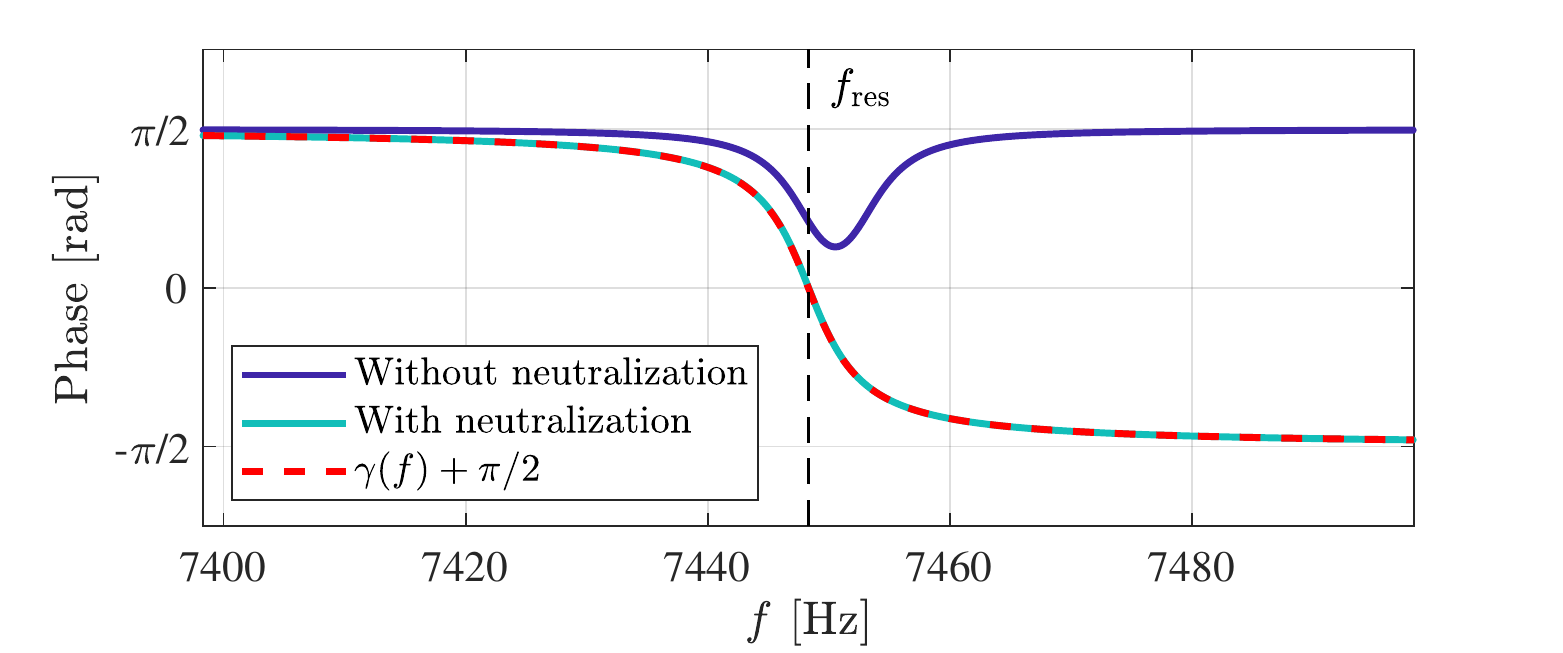}
		\caption{Phase of the sensor's admittance\\~}
		\label{fig:admittances_phase_of_f}
	\end{subfigure}
	\begin{subfigure}[t]{0.49\textwidth}
		\centering
		\includegraphics[width=1.0\linewidth]{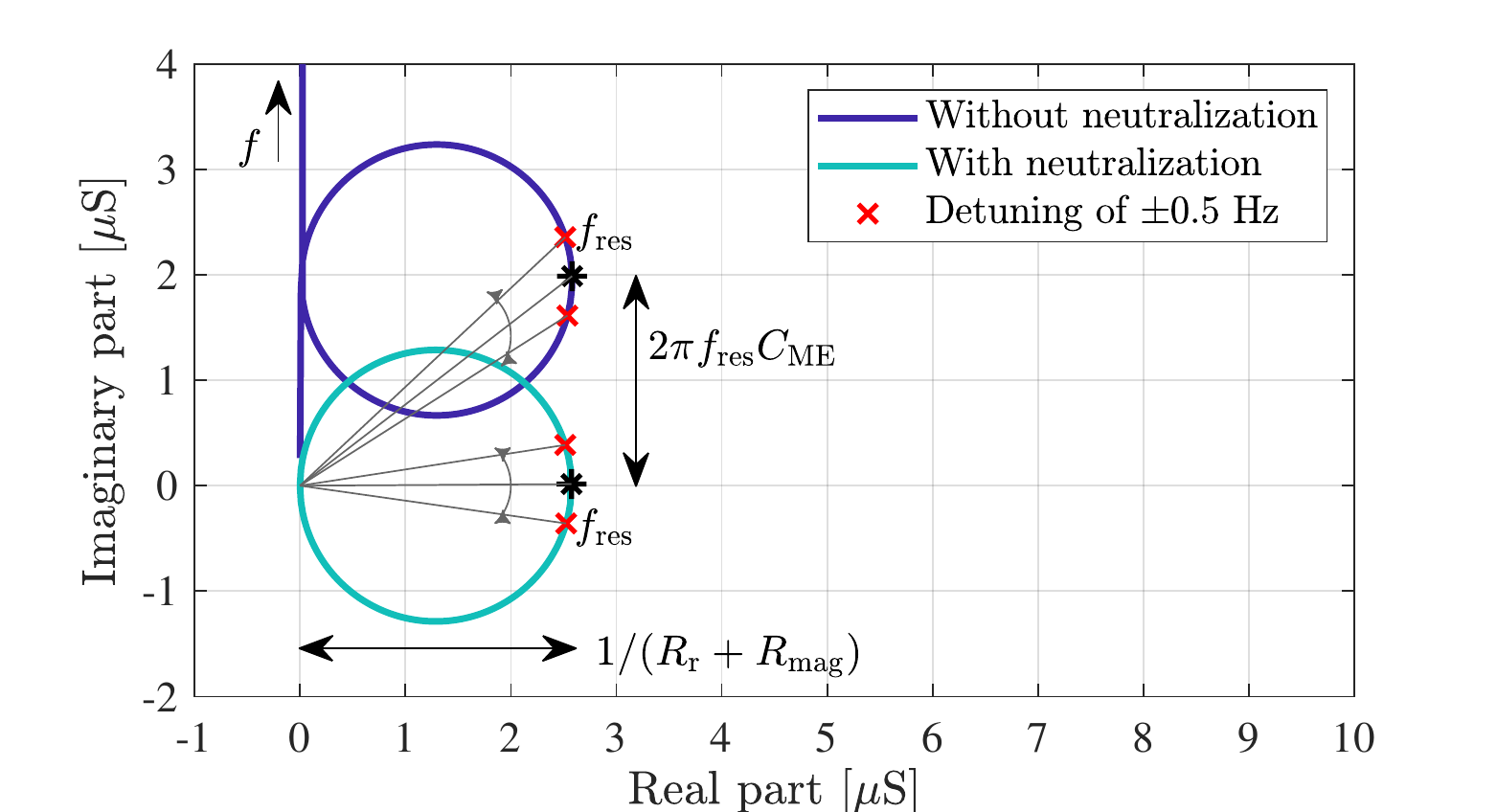}
		\caption{Nyquist plot of the sensor's admittance\\~}
		\label{fig:nyquist_plot}
	\end{subfigure}
	\begin{subfigure}[t]{0.49\textwidth}
		\centering
		\includegraphics[width=1.0\linewidth]{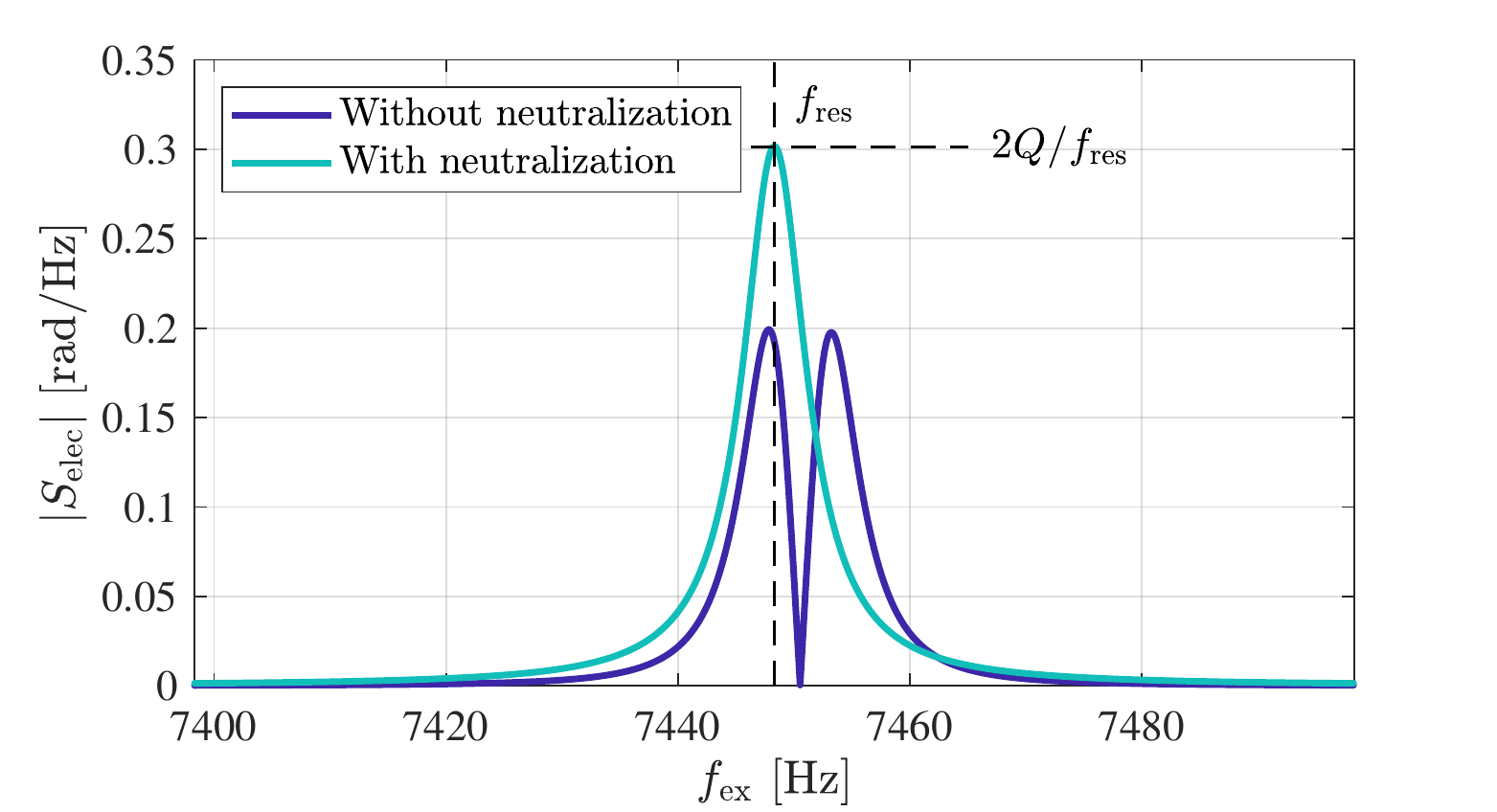}
		\caption{Electrical sensitivity\\~}
		\label{fig:Selec_of_f}
	\end{subfigure}
	\caption{Measured (stars) and calculated (solid lines) trajectories of magnitude (\subref{fig:admittances_magnitude_of_f}) and phase (\subref{fig:admittances_phase_of_f}) of the sensor's admittance $Y_{\mathrm{sensor}}$. Due to the parasitic influence of the sensor's static capacitance $C_{\mathrm{ME}}$ both line shapes are distorted, i.e. the locus curve (\subref{fig:nyquist_plot}) is shifted to higher imaginary parts. Neutralizing this effect not only leads to symmetric line shapes but also to an increase in the electrical sensitivity (\subref{fig:Selec_of_f}). The measurements have been conducted in magnetic saturation and for ${\hat{V}_{\mathrm{ex}} = \SI{1}{mV}}$.}
	\label{fig:admittances_of_f_nyquist_plot_Selec_of_f}
\end{figure*}

For sensor operation, i.e. for the reconstruction of a magnetic measurement signal $B_{\mathrm{x}}(t)$ due to an induced detuning of the resonator an electrical readout system as depicted in Fig.~\ref{fig:sensor_system} is utilized. The basic principle is based on a resonant excitation of the sensor with a voltage signal ${v_{\mathrm{ex}}(t) = \hat{V}_{\mathrm{ex}} \cos(2 \pi f_{\mathrm{ex}} t)}$ with ${f_{\mathrm{ex}} = f_{\mathrm{res}}}$ leading to a magnetically modulated current through the sensor $i_{\mathrm{sensor}}(t)$ that, in turn, is transformed into a proportional voltage signal $v_{\mathrm{sensor}}(t)$ utilizing a transimpedance amplifier and subsequent phase demodulation. For all measurements a low-noise JFET charge amplifier \cite{Dur17d} with a feedback capacitance of ${C_{\mathrm{f}} = \SI{30}{pF}}$ and a feedback resistance of ${R_{\mathrm{f}} = \SI{5}{G}\Omega}$ is utilized whose transimpedance in the vicinity of $f_{\mathrm{res}}$, i.e. far above the amplifier's lower cutoff frequency ${(2 \pi R_{\mathrm{f}} C_{\mathrm{f}})^{-1} \approx \SI{1}{Hz}}$, is given by
\begin{align}
	T(f) = \frac{V_{\mathrm{sensor}}(f)}{I_{\mathrm{sensor}}(f)} = -\frac{1}{j 2 \pi f C_{\mathrm{f}}}.
\end{align}
However, as already mentioned above, this type of electromechanical sensor has an additional static capacitance $C_{\mathrm{ME}}$ due to its electrodes that, with regard to the electrical equivalent circuit, appears in parallel to the series resonant circuit with the impedance $Z_{\mathrm{r}}$ leading to an overall admittance of the electromechanical sensor of
\begin{align}
	Y_{\mathrm{sensor}} = \frac{1}{Z_{\mathrm{r}}} + \frac{1}{R_{\mathrm{ME}}} + j \omega C_{\mathrm{ME}}.
\end{align}
As a consequence of the additional static capacitance $C_{\mathrm{ME}}$, a parallel resonance (also referred to as antiresonance) with a frequency ${f_{\mathrm{ares}} = f_{\mathrm{res}} \sqrt{1 + C_{\mathrm{r}}/C_{\mathrm{ME}}}}$ slightly above $f_{\mathrm{res}}$ \cite[p. 204]{Jon13} appears which distorts the line shape of both the magnitude (Fig.~\ref{fig:admittances_magnitude_of_f}) and the phase (Fig.~\ref{fig:admittances_phase_of_f}) of $Y_{\mathrm{sensor}}$ compared to that of a bare resonator described by $G(f)$ (Eq.~\eqref{eq:Gf}). As a result, the electrical sensitivity resulting by differentiation of the admittance's phase is smaller than stated in Eq.~\eqref{eq:Selec} (Fig.~\ref{fig:Selec_of_f}). The Nyquist plot in Fig.~\ref{fig:nyquist_plot} illustrates that effect by means of the phasor of the excitation signal, i.e. the carrier signal (gray lines), and by the highlighted resonance frequencies (red crosses) of the slightly detuned resonator. Although the effect is not as pronounced for the sensor under investigation due to its comparatively low static capacitance of only ${C_{\mathrm{ME}} = \SI{44}{pF}}$, one can easily imagine that the larger the sensor's static capacitance (shift of the locus curve to the top), the lower the resulting phase modulation. This is the reason why various methods for neutralizing the sensor's static capacitance have been reported \cite{Gro00,Arn08,Hu16,Fow17}, especially for large capacitance sensors. On the contrary, for this contribution, a neutralization is performed for symmetrizing the sensor's behavior. Thus, the admittance can be described by the frequency response of a simple harmonic oscillator $G(f)$ (compare Fig.~\ref{fig:admittances_magnitude_of_f} and Fig.~\ref{fig:admittances_phase_of_f}) which simplifies the noise considerations in the following section.

In the actual system as depicted in Fig.~\ref{fig:sensor_system} the neutralization is perceived by a second branch that contains a trimming capacitor with a capacitance of ${C_{\mathrm{n}} \approx C_{\mathrm{ME}}}$ and an identical charge amplifier as in the sensor branch. When neglecting the influence of the dielectric losses ($R_{\mathrm{ME}}$) on the sensor signal $v_{\mathrm{sensor}}(t)$ (see above) its amplitude spectrum yields
\begin{align}
	V_{\mathrm{sensor}}(f) &= T I_{\mathrm{sensor}}(f) = T V_{\mathrm{ex}} Y_{\mathrm{sensor}}(f)\\
	&= T V_{\mathrm{ex}} \left( \frac{1}{Z_{\mathrm{r}}(f)} + j 2 \pi f C_{\mathrm{ME}} \right).
\end{align}
Similarly, the amplitude spectrum at the output of the second charge amplifier is given by
\begin{align}
	V_{\mathrm{n}}(f) &= T I_{\mathrm{n}}(f) = T V_{\mathrm{ex}} j 2 \pi f C_{\mathrm{ME}},
\end{align}
thus resulting in an amplitude spectrum of the differential signal $v_{\mathrm{d}}(t)$
\begin{align}
	V_{\mathrm{d}}(f) &= V_{\mathrm{sensor}}(f) - V_{\mathrm{n}}(f) = \frac{T V_{\mathrm{ex}}}{Z_{\mathrm{r}}(f)}
\end{align}
in which the parasitic influence of the static capacitance $C_{\mathrm{ME}}$ is suppressed. For a resonant excitation ${Z_{\mathrm{r}}(f_{\mathrm{res}})}$ is purely ohmic. Consequently, the differential signal's amplitude can be written as (compare Fig.~\ref{fig:admittances_magnitude_of_f})
\begin{align}
	\hat{V}_{\mathrm{d}} = \sqrt{2}~|V_{\mathrm{d}}(f_{\mathrm{res}})| &= \frac{\hat{V}_{\mathrm{ex}} |T(f_{\mathrm{res}})| |G(f_{\mathrm{res}})|}{Q (R_{\mathrm{r}}+R_{\mathrm{mag}})}\\
	&= \frac{\hat{V}_{\mathrm{ex}} |T(f_{\mathrm{res}})|}{R_{\mathrm{r}}+R_{\mathrm{mag}}}.
	\label{eq:Vdpeak}
\end{align}
When neglecting the amplitude modulation and the static phase delays due to the sensor and the amplifiers, the associated time domain signal can be written as
\begin{align}
	v_{\mathrm{d}}(t) = \hat{V}_{\mathrm{d}}~\cos\left( 2 \pi f_{\mathrm{res}} t + S_{\mathrm{PM}} B_{\mathrm{x}}(t) + \varphi(t) \right)
	\label{eq:vdt}
\end{align}
which contains the phase modulation with the phase sensitivity $S_{\mathrm{PM}}$ (Eq.~\eqref{eq:SPM}) and phase fluctuations $\varphi(t)$ due to the sensor and the electronics that are analyzed in more detail in the following section.

By means of a quadrature detector the phase demodulation is performed in the digital domain. For ${f_{\mathrm{ex}} = f_{\mathrm{res}}}$ the output signal is then equal to 
\begin{align}
	s_{\mathrm{out}}(t) = S_{\mathrm{PM}} B_{\mathrm{x}}(t) + \varphi(t).
\end{align}
For all measurements in this paper, a high-resolution analog-to-digital (A/D) and digital-to-analog (D/A) converter, respectively, of type \textit{Fireface UFX} from \textit{RME} running at a sampling rate of \SI{32}{kHz} has been used for digitizing $v_{\mathrm{sensor}}(t)$ and $v_{\mathrm{n}}(t)$ and for generating the excitation signal $v_{\mathrm{ex}}(t)$. The digital low-pass filters (LPF) in the quadrature detector are third-order Butterworth filters with \SI{-3}{dB} cutoff frequencies of \SI{3}{kHz}.

\section{Phase Noise Analysis}
\label{sec:phase_noise_analysis}
\subsection{Thermal-mechanical and thermal-electrical noise}
\label{subsec:phase_noise_analysis_thermal_mechanical_and_thermal_electrical_noise}

\begin{figure}[t]
	\centering
	\includegraphics[width=0.5\textwidth]{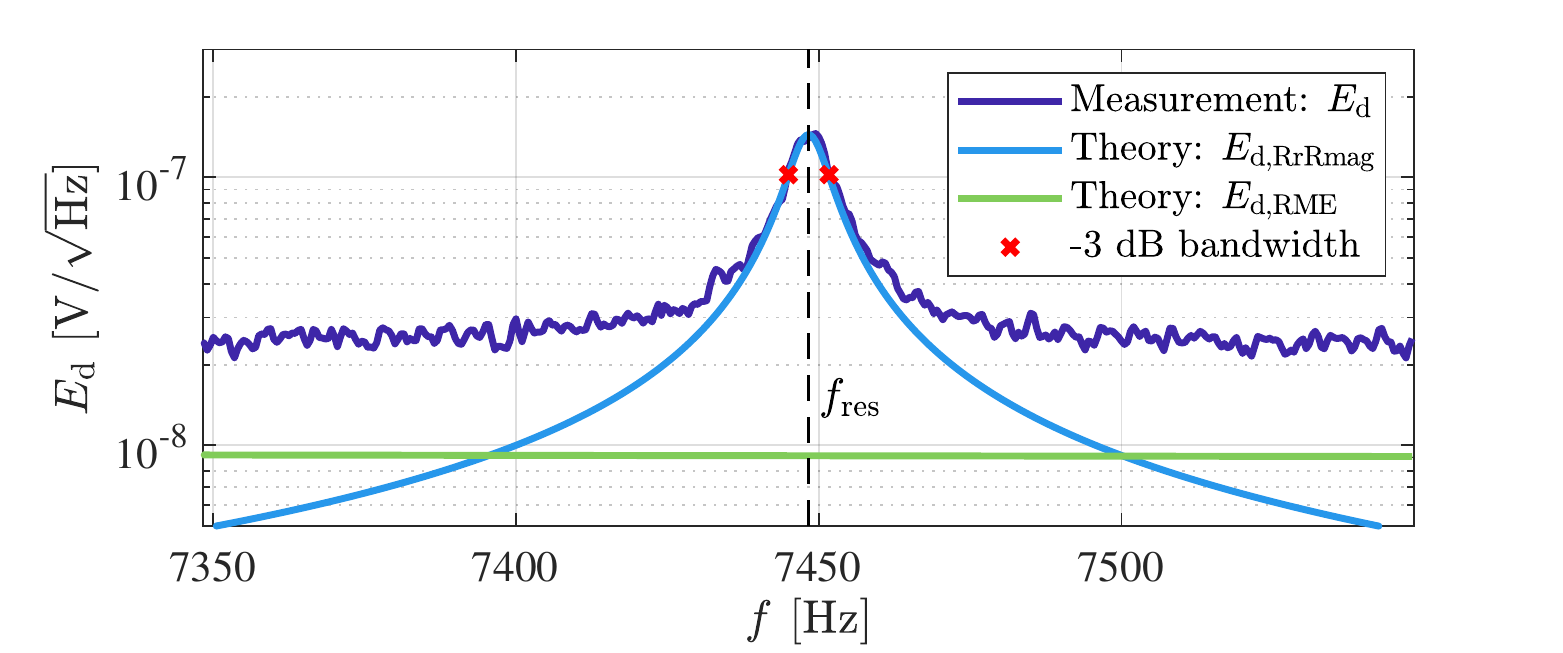}
	\caption{Measured noise at the output of the subtractor in comparison to the theoretical expectations according to Eq.~\eqref{eq:EdRrRmag} and \eqref{eq:EdRME}. In the vicinity of the resonance frequency the overall noise is dominated by the sensor-intrinsic thermal-mechanical noise $E_{\mathrm{d,RrRmag}}$. Far beyond the sensor's \SI{-3}{dB} bandwidth the measured noise is approximately frequency independent but higher than predicted ($E_{\mathrm{d,RME}}$) because of further contributions of the system electronics. The measurement has been conducted in magnetic saturation and for ${\hat{V}_{\mathrm{ex}} = 0}$.}
	\label{fig:Ed_sat_short}
\end{figure}

As discussed in Sec.~\ref{subsec:magnetoelastic_sensor_system_electrical_equivalent_circuit_and_loss_mechanisms}, an electromechanical cantilever exhibits several loss mechanisms that, in the electrical equivalent circuit, are covered by two resistors with the resistances ${R_{\mathrm{r}}+R_{\mathrm{mag}}}$ and $R_{\mathrm{ME}}$. In previous studies \cite{Dur17a,Dur17d} it has already been shown that both the related thermal-mechanical noise of the resonant structure $E_{\mathrm{d,RrRmag}}$ as well as the thermal-electrical noise $E_{\mathrm{d,RME}}$ of the dielectric material can accurately be predicted. Adjusting the previously published expressions to the readout structure as depicted in Fig.~\ref{fig:sensor_system}, the amount of the sensor's thermal voltage noise at the output of the subtractor can be calculated by
\begin{align}
	E_{\mathrm{d,RrRmag}}(f) &= \left|\frac{T(f)}{Z_{\mathrm{r}}(f)}\right| \sqrt{4 k_{\mathrm{B}} T_0 (R_{\mathrm{r}}+R_{\mathrm{mag}})}\\
	&= \frac{|T(f)| |G(f)| \sqrt{4 k_{\mathrm{B}} T_0}}{Q \sqrt{R_{\mathrm{r}} + R_{\mathrm{mag}}}}
	\label{eq:EdRrRmag}
\end{align}
and
\begin{align}
	E_{\mathrm{d,RME}}(f) &= \frac{|T(f)|}{R_{\mathrm{ME}}(f)} \sqrt{4 k_{\mathrm{B}} T_0 R_{\mathrm{ME}}(f)}\\
	&= \frac{|T(f)| \sqrt{4 k_{\mathrm{B}} T_0}}{\sqrt{R_{\mathrm{ME}}(f)}}
	\label{eq:EdRME}
\end{align}
where ${k_{\mathrm{B}} \approx 1.381 \cdot 10^{-23}~\mathrm{J/K}}$ is the Boltzmann constant and ${T_0 = \SI{290}{K}}$ the room temperature. A noise measurement without any external excitation of the sensor (${\hat{V}_{\mathrm{ex}} = 0}$) and in comparison to the theoretical expectations is depicted in Fig.~\ref{fig:Ed_sat_short}. Due to the sensor's resonant behavior the thermal-mechanical noise $E_{\mathrm{d,RrRmag}}$ is weighted by $G(f)$ and perfectly agrees with the measurement in the vicinity of the resonance frequency. Far beyond the sensor's \SI{-3}{dB} bandwidth the measured noise is approximately frequency independent but higher than predicted ($E_{\mathrm{d,RME}}$) because of further contributions due to the system electronics, i.e. that of the two charge amplifiers and the D/A and A/D converters, respectively.

\subsection{Relation between voltage noise density and phase noise}
\label{subsec:phase_noise_analysis_relation_between_voltage_noise_density_and_phase_noise}

Various and statistically independent voltage noise densities due to the sensor and the system electronics ($E_{\mathrm{d,system}}$) add up at the output of the subtractor
\begin{align}
	E_{\mathrm{d}}^2 = E_{\mathrm{d,RrRmag}}^2 + E_{\mathrm{d,RME}}^2 + E_{\mathrm{d,system}}^2.
\end{align}
To determine their relation to the power spectral density $S_\varphi(f)$ of the random phase fluctuations $\varphi(t)$, Eq.~\eqref{eq:vdt} is written as
\begin{align}
	v_{\mathrm{d}}(t) = \hat{V}_{\mathrm{d}}~\cos\left( 2 \pi f_{\mathrm{res}} t + \hat{\varphi}(f_{\mathrm{x}}) \cos(2 \pi f_{\mathrm{x}} t) \right)
	\label{eq:vdt_noise1}
\end{align}
in which one noise component with the modulation index $\hat{\varphi}$ at $f_{\mathrm{x}}$ represents other spectral components that can be taken into account by linear superposition. In addition, for this noise consideration the measurement signal $B_{\mathrm{x}}(t)$ is assumed to be zero. Based on basic trigonometric identities Eq.~\eqref{eq:vdt_noise1} can be rearranged into 
\begin{align}
	v_{\mathrm{d}}(t) &= \hat{V}_{\mathrm{d}} \bigg[ \cos(2 \pi f_{\mathrm{res}} t) - \frac{\hat{\varphi}(f_{\mathrm{x}})}{2} \big[ \sin(2 \pi (f_{\mathrm{res}}-f_{\mathrm{x}}) t)\notag\\
	&\hspace{2cm}+ \sin(2 \pi (f_{\mathrm{res}}+f_{\mathrm{x}}) t) \big] \bigg]
\end{align}
revealing the typical structure of a narrow band small signal phase modulated signal with a carrier at $f_{\mathrm{res}}$ and two symmetrical sidebands at ${f_{\mathrm{res}} \pm f_{\mathrm{x}}}$. Following the concept of noise sidebands \cite[pp. 243]{Lan84} the carrier-to-noise sideband ratio
\begin{align}
	\frac{\hat{V}_{\mathrm{d}}}{\hat{V}_{\mathrm{d}} \frac{\hat{\varphi}(f_{\mathrm{x}})}{2}} = \frac{2}{\hat{\varphi}(f_{\mathrm{x}})} = \frac{\hat{V}_{\mathrm{d}}}{E_{\mathrm{d}}(f_{\mathrm{res}} \pm f_{\mathrm{x}})~\sqrt{2} \sqrt{\Delta f}}
	\label{eq:ctnsr}
\end{align}
is equal to the carrier-to-voltage noise ratio if a symmetrical noise distribution (${E_{\mathrm{d}}(f_{\mathrm{res}} - f_{\mathrm{x}}) = E_{\mathrm{d}}(f_{\mathrm{res}} + f_{\mathrm{x}})}$) around the resonance frequency is assumed. The additional term $\sqrt{\Delta f}$ transforms the voltage noise density into an effective voltage noise in the bandwidth $\Delta f$. From Eq.~\eqref{eq:ctnsr} the phase modulation index 
\begin{align}
	\hat{\varphi}(f_{\mathrm{x}}) = \frac{2~E_{\mathrm{d}}(f_{\mathrm{res}} \pm f_{\mathrm{x}})~\sqrt{2} \sqrt{\Delta f}}{\hat{V}_{\mathrm{d}}}
\end{align}
can be deduced which directly yields the power spectral density 
\begin{align}
	S_{\varphi}(f_{\mathrm{x}}) = \left(\frac{2~E_{\mathrm{d}}(f_{\mathrm{res}} \pm f_{\mathrm{x}})}{\hat{V}_{\mathrm{d}}}\right)^2
\end{align}
of the random phase fluctuations $\varphi(t)$ in units of ${\mathrm{rad}^2/\mathrm{Hz}}$. For voltage noise distributed asymmetrically around the resonance frequency (${E_{\mathrm{d}}(f_{\mathrm{res}} - f_{\mathrm{x}}) \neq E_{\mathrm{d}}(f_{\mathrm{res}} + f_{\mathrm{x}})}$) the more generally valid power spectral density is given by
\begin{align}
	S_{\varphi}(f_{\mathrm{x}}) = \left(\frac{E_{\mathrm{d}}(f_{\mathrm{res}} - f_{\mathrm{x}}) + E_{\mathrm{d}}(f_{\mathrm{res}} + f_{\mathrm{x}})}{\hat{V}_{\mathrm{d}}}\right)^2.
\end{align}
However, for the frequency range in the vicinity of $f_{\mathrm{res}}$ 
\begin{align}
	|T(f_{\mathrm{res}})| \approx \frac{1}{2} \left( |T(f_{\mathrm{res}} - f_{\mathrm{x}})| + |T(f_{\mathrm{res}} + f_{\mathrm{x}})| \right)
\end{align}
and
\begin{align}
	R_{\mathrm{ME}}(f_{\mathrm{res}}) \approx \frac{1}{2} \left( R_{\mathrm{ME}}(f_{\mathrm{res}} - f_{\mathrm{x}}) + R_{\mathrm{ME}}(f_{\mathrm{res}} + f_{\mathrm{x}}) \right)
\end{align}
are generally good approximations. Because of the neutralization $E_{\mathrm{d,RrRmag}}(f)$ is also symmetric around $f_{\mathrm{res}}$, thus leading to expressions for the power spectral densities of random phase fluctuations due to thermal-mechanical noise
\begin{align}
	S_{\varphi,\mathrm{RrRmag}}(f_{\mathrm{x}}) &= \left(\frac{2~|T(f_{\mathrm{res}}) S_{\mathrm{dyn}}(f_{\mathrm{x}})| \sqrt{4 k_{\mathrm{B}} T_0}}{\hat{V}_{\mathrm{d}} \sqrt{R_{\mathrm{r}} + R_{\mathrm{mag}}}}\right)^2\\
	&\hspace{-1.5cm} \overset{\text{Eq.~\eqref{eq:Vdpeak}}}{=} \left(\frac{2~|S_{\mathrm{dyn}}(f_{\mathrm{x}})| \sqrt{4 k_{\mathrm{B}} T_0 (R_{\mathrm{r}} + R_{\mathrm{mag}})}}{\hat{V}_{\mathrm{ex}}}\right)^2
	\label{eq:SvarphiRrRmagfx}
\end{align}
and due to thermal-electrical noise of the loss in the dielectric material
\begin{align}
	S_{\varphi,\mathrm{RME}} &= \left(\frac{2~|T(f_{\mathrm{res}})| \sqrt{4 k_{\mathrm{B}} T_0}}{\hat{V}_{\mathrm{d}} \sqrt{R_{\mathrm{ME}}(f_{\mathrm{res}})}}\right)^2\\
	&\overset{\text{Eq.~\eqref{eq:Vdpeak}}}{=} \left(\frac{2~\sqrt{4 k_{\mathrm{B}} T_0}~(R_{\mathrm{r}}+R_{\mathrm{mag}})}{\hat{V}_{\mathrm{ex}} \sqrt{R_{\mathrm{ME}}(f_{\mathrm{res}})}}\right)^2.
	\label{eq:SvarphiRMEfx}
\end{align}
Obviously, the thermal-electrical noise $E_{\mathrm{d,RME}}$ leads to additive white phase noise that decreases with the excitation amplitude since ${\hat{V}_{\mathrm{d}} \propto \hat{V}_{\mathrm{ex}}}$. The phase noise, which is caused by the thermal-mechanical noise $E_{\mathrm{d,RrRmag}}$, also decreases with $\hat{V}_{\mathrm{ex}}$. However, due to the influence of the resonator this additive phase noise decreases with \SI{20}{dB/decade} for frequencies $f_{\mathrm{x}}$ above the cutoff frequency ${f_{\mathrm{c}} = f_{\mathrm{res}}/(2Q)}$. In contrast to this, thermal-mechanical noise also leads to white phase noise for frequencies well below the cutoff frequency (${f_{\mathrm{x}} \ll f_{\mathrm{c}}}$).

\subsection{Phase noise measurements}
\label{subsec:phase_noise_analysis_measurements_phase noise_measurements}

\begin{figure*}[t]
	\captionsetup[subfigure]{justification=centering}
	\centering
	\begin{subfigure}[t]{0.49\textwidth}
		\centering
		\includegraphics[width=1.0\linewidth]{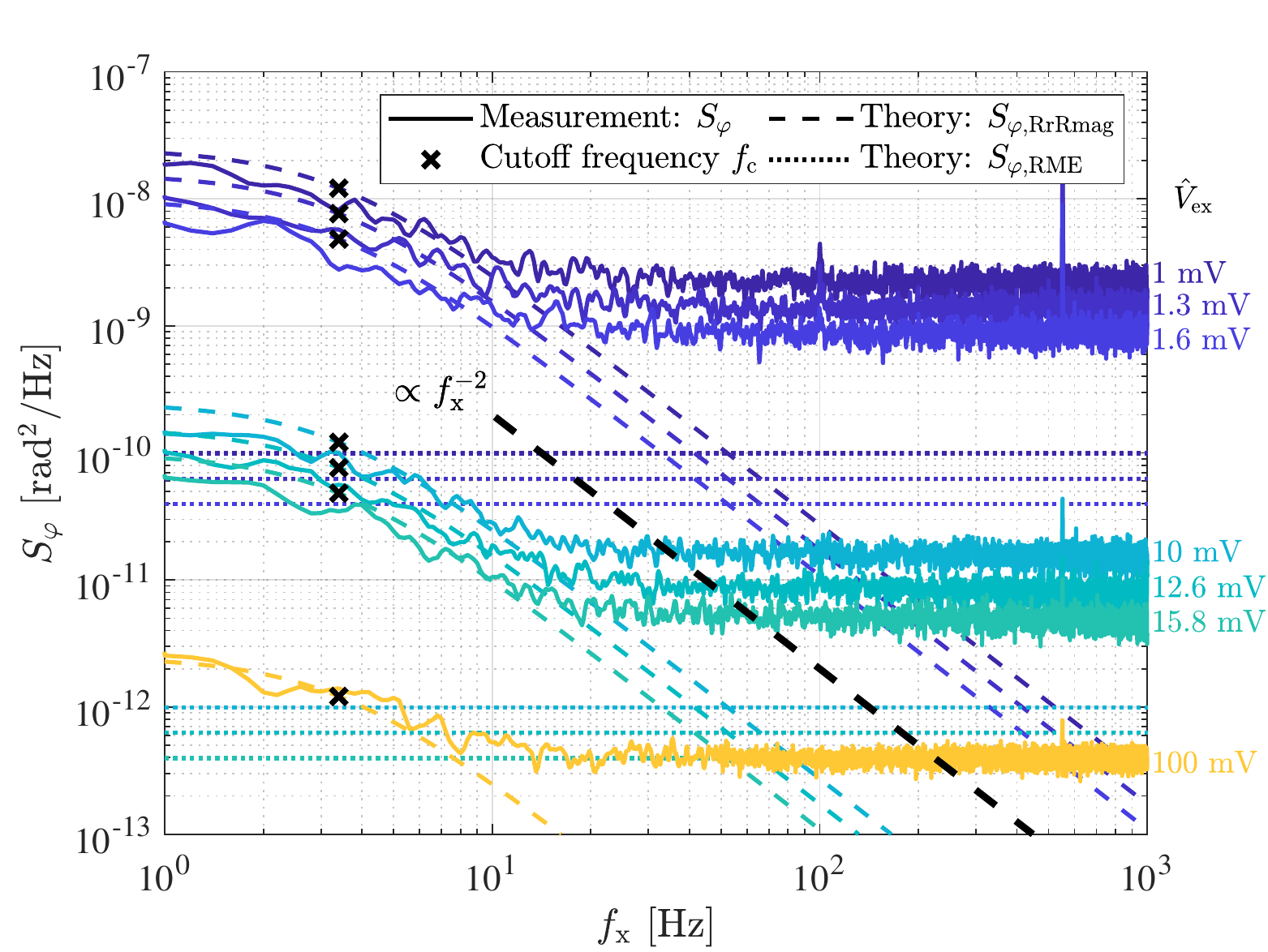}
		\caption{Sensor in magnetic saturation\\~}
		\label{fig:phase_noise_sweep_at_bsat}
	\end{subfigure}
	\begin{subfigure}[t]{0.49\textwidth}
		\centering
		\includegraphics[width=1.0\linewidth]{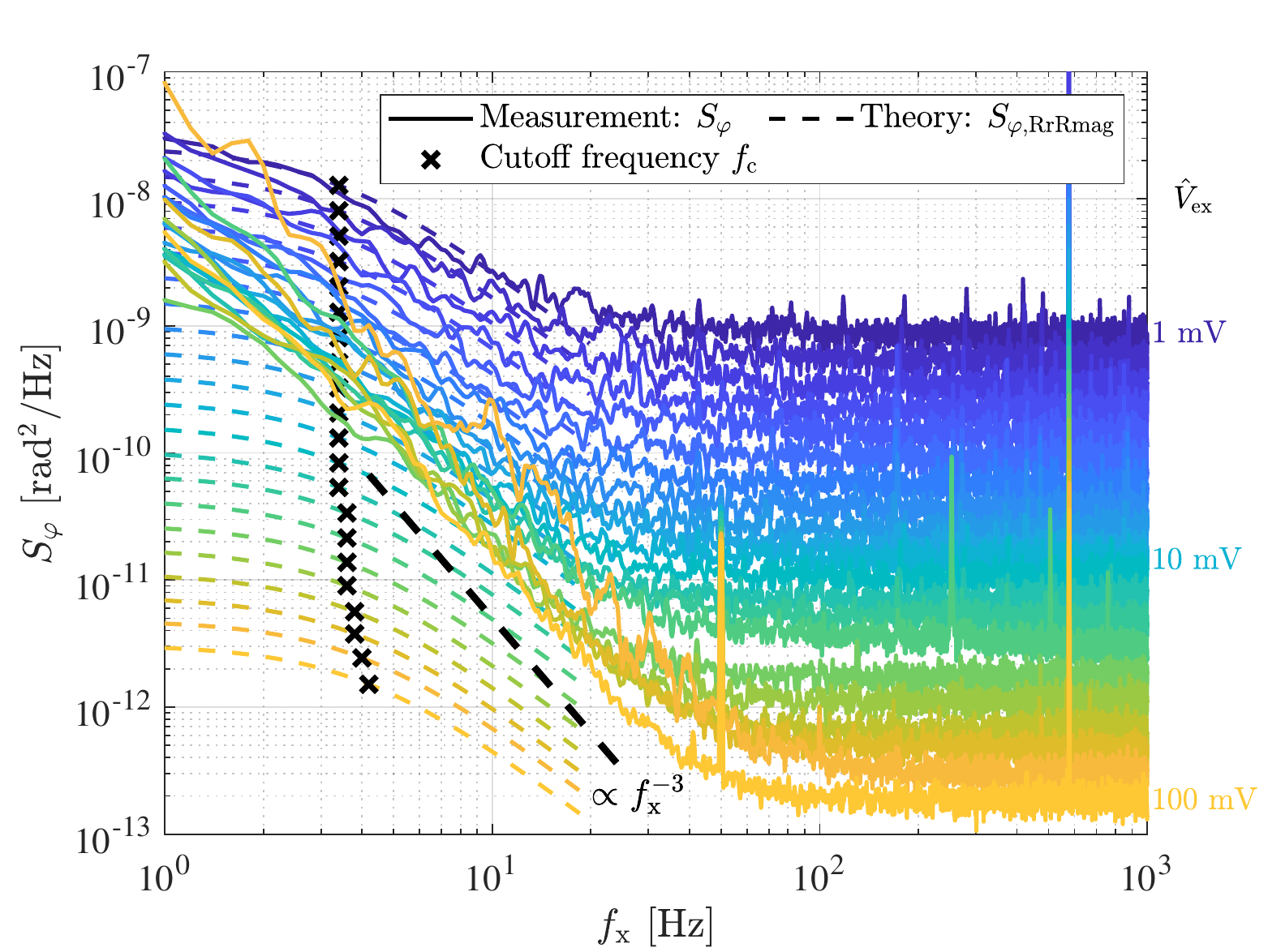}
		\caption{Sensor in magnetic operating point ${B_{\mathrm{bias}} = \SI{0.65}{mT}}$\\~}
		\label{fig:phase_noise_sweep_at_bbiasmax}
	\end{subfigure}
	\caption{Power spectral densities of the measured random phase fluctuations $\varphi(t)$ at the sensor system's output for the sensor in magnetic saturation (\subref{fig:phase_noise_sweep_at_bsat}) and for the sensor in its magnetic operating point (\subref{fig:phase_noise_sweep_at_bbiasmax}) for various amplitudes of the excitation signal. For the magnetically saturated sensor, the additive phase noise due to thermal-mechanical (Eq.~\eqref{eq:SvarphiRrRmagfx}, dashed lines) and due to thermal-electrical noise (Eq.~\eqref{eq:SvarphiRMEfx}, dotted lines) decreases with the excitation amplitude $\hat{V}_{\mathrm{ex}}$. For the sensor in its magnetic operating point low-frequency parametric phase noise occurs whose underlying physical noise process must exhibit a ${f_{\mathrm{x}}^{-1}}$ characteristic because an amount of ${f_{\mathrm{x}}^{-2}}$ is attributed to the influence of the resonator. The slightly different levels of white phase noise floors between the measurements in (\subref{fig:phase_noise_sweep_at_bsat}) and (\subref{fig:phase_noise_sweep_at_bbiasmax}) are probably due to different states of charge of the battery supplying the amplifier, thus leading to different drain currents in the amplifier's discrete JFET front-end \cite{Dur17d}.}
	\label{fig:phase_noise_sweep_at_bsat_and_bbiasmax}
\end{figure*}

\begin{figure*}[t]
	\captionsetup[subfigure]{justification=centering}
	\centering
	\begin{subfigure}[t]{0.475\textwidth}
		\centering
		\includegraphics[width=1.0\linewidth]{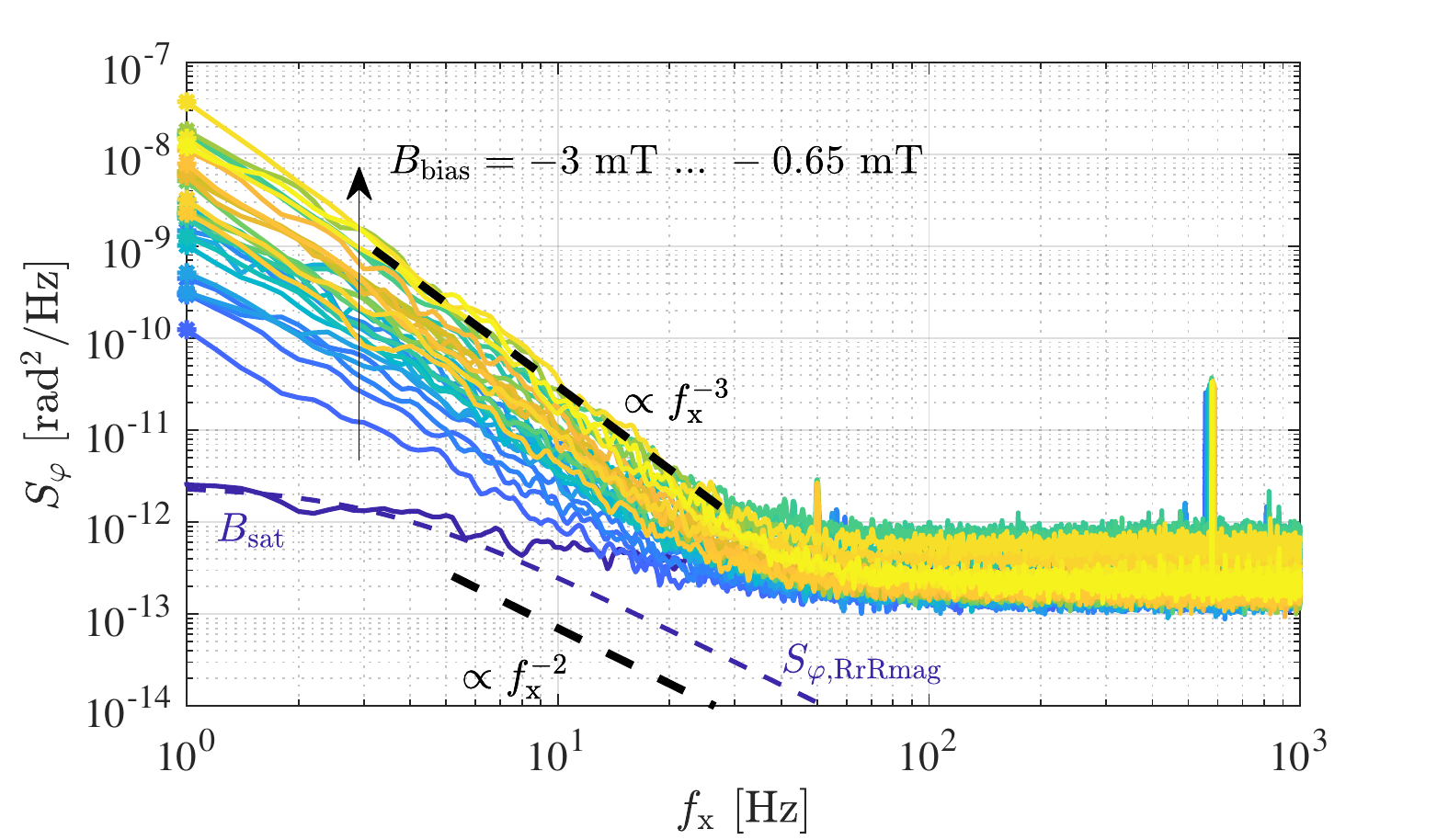}
		\caption{Phase noise for various magnetic bias flux densities in the frequency domain\\~}
		\label{fig:phase_noise_bbias_sweep_f}
	\end{subfigure}
	\begin{subfigure}[t]{0.475\textwidth}
		\centering
		\includegraphics[width=1.0\linewidth]{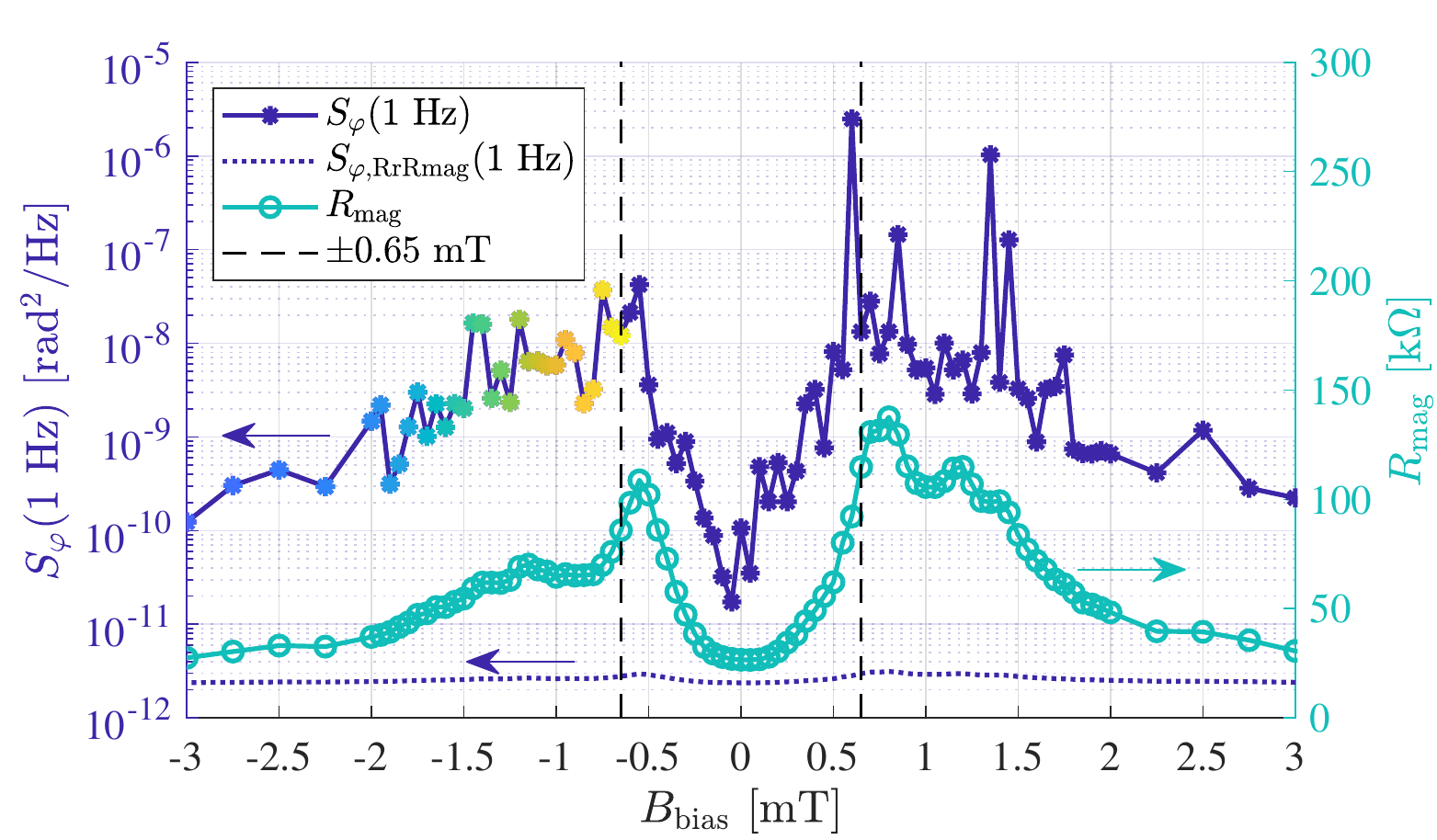}
		\caption{Phase noise at a frequency of \SI{1}{Hz} and magnetic losses as a function of the magnetic bias flux density\\~}
		\label{fig:phase_noise_bbias_sweep_bbias}
	\end{subfigure}
	\caption{Power spectral densities of the measured random phase fluctuations $\varphi(t)$ at the sensor system's output for a constant excitation amplitude of ${\hat{V}_{\mathrm{ex}} = \SI{100}{mV}}$ as a function of the magnetic bias flux density $B_{\mathrm{bias}}$ measured increasingly from negative to positive magnetic saturation. The magnetically induced phase noise significantly changes with $B_{\mathrm{bias}}$ while the slope stays constant as far as the sensor is not in magnetic saturation (\subref{fig:phase_noise_bbias_sweep_f}). Obviously, the measured phase noise is directly linked to the magnetic losses, represented by $R_{\mathrm{mag}}$, and for this excitation amplitude always higher than the phase noise due to the sensor's thermal-mechanical noise $S_{\varphi,\mathrm{RrRmag}}$ (\subref{fig:phase_noise_bbias_sweep_bbias}).}
	\label{fig:phase_noise_bbias_sweep}
\end{figure*}

With the system described above (Fig.~\ref{fig:sensor_system}), several series of noise measurements were performed to analyze the sensor's phase noise behavior. Without any additional magnetic measurement signal (${B_{\mathrm{x}}(t) = 0}$), the system's output signal $s_{\mathrm{out}}(t)$ is then equal to the random phase fluctuations $\varphi(t)$ which were transformed to the frequency domain based on Welch's method \cite{Wel67}, thus leading to the power spectral densities of random phase fluctuations $S_{\varphi}$. In the following, $S_{\varphi}(f_{\mathrm{x}})$ is given as a function of the frequency $f_{\mathrm{x}}$ to clarify that this phase noise is effective in the same frequency range as the measurement signal. Formally it would be just as correct to use $f$ here.

For all measurements the sensor was placed inside an ultra-high magnetic field shielding mu-metal cylinder (Aaronia AG ZG1) which, in turn, is placed inside a vibrationally decoupled box with lined absorbers against airborne sound. In addition, the whole box is coated with a copper fleece, which shields electrical fields. Furthermore, the sensor is surrounded by two solenoids inside the mu-metal that are used for generating both the magnetic bias flux density $B_{\mathrm{bias}}$ as well as the magnetic measurement signal $B_{\mathrm{x}}(t)$. The latter is generated with a commercially available current source (Keithley 6221). However, no commercially available power source was suitable for generating comparatively large bias fields in the millitesla range while keeping the resulting low-frequency noise well below ${100~\mathrm{pT}/\sqrt{\mathrm{Hz}}}$. Therefore, an in-house built setup based on several batteries (capacity $> \SI{100}{Ah}$) and a stepper motor controlled potentiometer in series to the coil was used.

In a first series of noise measurements the sensor was magnetically saturated by means of a strong permanent magnet generating a magnetic bias flux density of ${B_{\mathrm{bias}} \approx \SI{20}{mT}}$ which is distinctly higher than the sensor's saturation flux density \SI{< 5}{mT} (compare e.g. Fig.~\ref{fig:fres_of_Bbias_for_Vex_10_mV}). At the same time the amplitude $\hat{V}_{\mathrm{ex}}$ of the electrical excitation signal $V_{\mathrm{ex}}(t)$ was increased incrementally. For several values of $\hat{V}_{\mathrm{ex}}$, the measured phase noise is depicted in Fig.~\ref{fig:phase_noise_sweep_at_bsat} together with the phase noise contributions due to thermal-mechanical noise $S_{\varphi,\mathrm{RrRmag}}$ (Eq.~\eqref{eq:SvarphiRrRmagfx}, dashed lines) and due to thermal-electrical noise of the dielectric material $S_{\varphi,\mathrm{RME}}$ (Eq.~\eqref{eq:SvarphiRMEfx}, dotted lines). As expected, the measured noise spectra decrease with increasing excitation amplitudes, thus confirming the additive character of both thermal noise contributions. In addition, values of measured phase noise and calculated phase noise $S_{\varphi,\mathrm{RrRmag}}$ (dashed lines) agree well for the low frequency range in the range of the sensor's passband, indicated by the respective cutoff frequency (black crosses). Due to the impact of the mechanical resonator, the phase noise decreases with higher frequencies proportional to ${f_{\mathrm{x}}^{-2}}$ until the white phase noise floor beyond the sensor's passband is reached. According to previous calculations, the crossover frequency between frequency-dependent and white phase noise due to thermal-electrical noise of the dielectric material (${S_{\varphi,\mathrm{RrRmag}} = S_{\varphi,\mathrm{RME}}}$) is about \SI{50}{Hz}. However, as already discussed in the context of Fig.~\ref{fig:Ed_sat_short}, the overall noise floor outside the sensor's passband is dominated by noise contributions of the system electronics, thus leading to a crossover frequency at about \SI{10}{Hz} for the system under investigation.

The same series of measurements was repeated for the sensor not being saturated anymore but brought into its magnetic operating point of ${B_{\mathrm{bias}} = \SI{0.65}{mT}}$ after saturating the sensor in negative direction and stepwise incrementing $B_{\mathrm{bias}}$. The measured phase noise acquired in this way is shown in Fig.~\ref{fig:phase_noise_sweep_at_bbiasmax}, again in comparison to the theoretical expectations (dashed lines) according to Eq.~\eqref{eq:SvarphiRrRmagfx} (phase noise due to thermal-mechanical noise $S_{\varphi,\mathrm{RrRmag}}$). The measured phase noise at low excitation amplitudes of about \SI{1}{mV} still corresponds to the contribution of the thermal-mechanical noise. However, for increasing amplitudes $\hat{V}_{\mathrm{ex}}$ the measured phase noise in the sensor's passband no longer decreases significantly as for the magnetically saturated case (Fig.~\ref{fig:phase_noise_sweep_at_bsat}). Thus, the noise contribution of the magnetic material leads to so-called parametric noise which is independent from the amplitude of the carrier signal \cite[p. 36]{Rub09}, at least if the noise process itself or material properties do not depend on the amplitude. Such a behavior is well-known from $1/f$ flicker phase noise, e.g. of amplifiers \cite{Bou12}. And indeed, considering the slope of ${f_{\mathrm{x}}^{-3}}$ with which the measured phase noise decreases, the underlying physical noise process must exhibit a ${f_{\mathrm{x}}^{-1}}$ characteristic because an amount of ${f_{\mathrm{x}}^{-2}}$ is attributed to the influence of the resonator.

To further verify the relationship between the sensor's magnetic state and the magnetically induced phase noise, measurements as a function of the magnetic bias flux density $B_{\mathrm{bias}}$ from negative to positive saturation were conducted (inverse measurement gives results mirrored on the axis of ordinates). As depicted by the corresponding power spectral densities of the random phase fluctuations in Fig.~\ref{fig:phase_noise_bbias_sweep_f}, the ${f_{\mathrm{x}}^{-1}}$ flicker phase noise and the ${f_{\mathrm{x}}^{-3}}$ phase noise at the sensor system's output, respectively, clearly depend on the sensor's magnetic state. In fact, as shown by the measured phase noise at a frequency of \SI{1}{Hz} in Fig.~\ref{fig:phase_noise_bbias_sweep_bbias}, the induced phase noise is unambiguously related to the magnetic losses, represented by $R_{\mathrm{mag}}$. For operating points with low losses, i.e. at which the magnetic sensitivity is low, e.g. near saturation and for ${B_{\mathrm{bias}} = 0}$, also the phase noise adopts lower values. In contrast, the phase noise is particularly high when the losses or the magnetic sensitivity is high (dashed lines at ${B_{\mathrm{bias}} = \SI{\pm0.65}{mT}}$). In investigations on magnetoresistive sensors, an identical behavior could be observed in the past \cite{Har93,Vee97}. These sensors also show largest noise for operating points of maximum sensitivity which was attributed to random fluctuations of the magnetization due to magnetic domain wall movements and rotations \cite{Vee97,Ing00}.

\subsection{Magnetically induced flicker phase noise}
\label{subsec:phase_noise_analysis_measurements_magnetically_induced_flicker_phase_noise}

Due to the significant relation to the magnetic losses it is obvious to describe the magnetically induced phase noise using the fluctuation-dissipation theorem. Based on that theorem, the power spectral density of random fluctuations of the magnetization $M$
\begin{align}
	S_{\mathrm{M}}(f_{\mathrm{x}}) = \frac{4 k_{\mathrm{B}} T_0}{2 \pi f_{\mathrm{x}}~V_{\mathrm{mag}}} \frac{\mu_{\mathrm{r,eff}}''}{\mu_0}
	\label{eq:SMfx}
\end{align}
with the physical dimension ${(\mathrm{A}/\mathrm{m})^2/\mathrm{Hz}}$ can be derived \cite{Dur93,Bri00} which can be referred to as flicker magnetization noise since the power density decreases with $1/f_{\mathrm{x}}$. This expression is typically given as a function of the imaginary part $\mu_{\mathrm{r}}''$ of the magnetic material's complex permeability ${\mu_{\mathrm{r}} = \mu_{\mathrm{r}}' -j \mu_{\mathrm{r}}''}$. In general, however, $\mu_{\mathrm{r}}''$ is also used to account for other losses, in particular eddy current losses, which in turn do not correspond with flicker noise but with frequency-independent white noise \cite{Lee08}. Therefore, in this paper, an effective complex permeability ${\mu_{\mathrm{r,eff}} = \mu_{\mathrm{r,eff}}' -j \mu_{\mathrm{r,eff}}''}$ is used to cover only for magnetic hysteresis losses corresponding with $1/f$ flicker noise. Furthermore, ${\mu_0 \approx 4 \pi \cdot 10^{-7}~\mathrm{V s}/(\mathrm{A m})}$ and $V_{\mathrm{mag}}$ denote the vacuum permeability and the volume of the magnetic material, respectively. Thus, fluctuations of the magnetization can be decreased by larger magnetic volumes, at least if the magnetic losses $\mu_{\mathrm{r,eff}}''$ do not rise proportionally with $V_{\mathrm{mag}}$. However, literature shows that volume and losses are generally not independent of each other \cite{Jay98}.

The expression for changes of the resonant sensor's phase response $\gamma(f)$ (Eq.~\ref{eq:gammaf}) at the resonance frequency $f_{\mathrm{res}}$ due to changes of the magnetization $M$
\begin{align}
	\frac{\partial \gamma(f_{\mathrm{res}})}{\partial M} = \frac{\partial \gamma(f)}{\partial f}\Bigg|_{f = f_{\mathrm{res}}} \frac{\partial f_{\mathrm{res}}}{\partial M}
\end{align}
can be factorized into two terms. The first term describes the changes of the sensor's phase response at $f_{\mathrm{res}}$ due to a detuning of the resonator. As discussed above, this term is equal to the electrical sensitivity $S_{\mathrm{elec}}$ (Eq.~\eqref{eq:Selec}). The second term covers the detuning of the resonator due to changes of the magnetization. With the magnetic susceptibility ${\chi = \partial M/\partial H = \mu_{\mathrm{r,eff}}'-1}$, with ${B = \mu_0 H}$, and with ${S_{\mathrm{mag}} = \partial f_{\mathrm{res}}/\partial B}$ (Eq.~\eqref{eq:Smag}) the second term
\begin{align}
	\frac{\partial f_{\mathrm{res}}}{\partial M} &= \frac{\partial f_{\mathrm{res}}}{\partial H} \frac{\partial H}{\partial M} = \frac{\partial f_{\mathrm{res}}}{\partial H} \frac{1}{\chi} = \frac{\partial f_{\mathrm{res}}}{\mu_0 \partial H} \frac{\mu_0}{\chi}\\
	&= S_{\mathrm{mag}} \frac{\mu_0}{\chi} \approx S_{\mathrm{mag}} \frac{\mu_0}{\mu_{\mathrm{r,eff}}'}
\end{align}
can be expressed as a function of the magnetic sensitivity $S_{\mathrm{mag}}$ and the real part $\mu_{\mathrm{r,eff}}'$ of the effective permeability. The approximation is generally valid for commonly utilized magnetic materials with high permeabilities (${\mu_{\mathrm{r,eff}}' \gg 1}$).

Using these relations, the power spectral density of random phase fluctuations due to random fluctuations of the magnetization yields
\begin{align}
	S_{\varphi,\mathrm{M}}(f_{\mathrm{x}}) &= S_{\mathrm{M}}(f_{\mathrm{x}}) \left|\frac{\partial \gamma(f_{\mathrm{res}})}{\partial M}\right|^2 |S_{\mathrm{dyn}}(f_{\mathrm{x}})|^2\\
	&\hspace{-0.5cm}= S_{\mathrm{M}}(f_{\mathrm{x}}) |S_{\mathrm{elec}}|^2 \left|S_{\mathrm{mag}} \frac{\mu_0}{\mu_{\mathrm{r,eff}}'}\right|^2 |S_{\mathrm{dyn}}(f_{\mathrm{x}})|^2
	\label{eq:SvarphifxM1}
\end{align}
in which the dynamic sensitivity $S_{\mathrm{dyn}}$ (Eq.~\eqref{eq:Sdynfx}) accounts for the additional decrease in phase noise with increasing frequency due to the resonator. With Eq.~\eqref{eq:SMfx} and the sensor's overall phase sensitivity $S_{\mathrm{PM}}$ (Eq.~\eqref{eq:SPM}) the expression further simplifies to 
\begin{align}
	S_{\varphi,\mathrm{M}}(f_{\mathrm{x}}) = \frac{4 k_{\mathrm{B}} T_0}{2 \pi f_{\mathrm{x}}~V_{\mathrm{mag}}}~|S_{\mathrm{PM}}(f_{\mathrm{x}})|^2~\frac{\mu_0 \mu_{\mathrm{r,eff}}''}{(\mu_{\mathrm{r,eff}}')^2},
	\label{eq:SvarphifxM2}
\end{align}
clarifying that the magnetically induced phase noise is proportional to the sensor's sensitivity.

\subsection{Limit of detection}
\label{subsec:phase_noise_analysis_measurements_limit_of_detection}

\begin{figure}[t]
	\centering
	\includegraphics[width=0.5\textwidth]{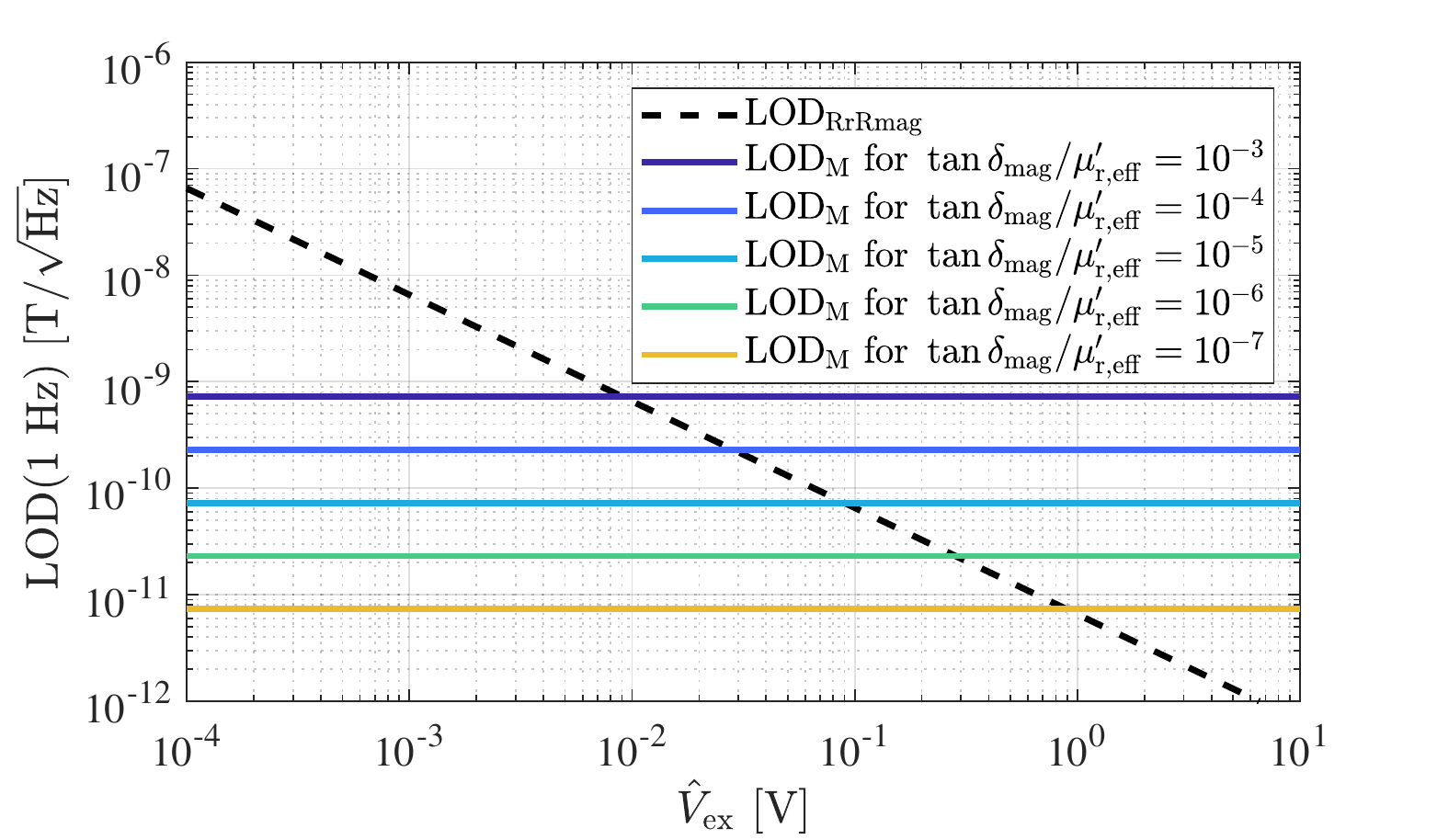}
	\caption{Calculated limit of detection (LOD) at a frequency of \SI{1}{Hz} and for typical sensor parameters (${f_{\mathrm{res}} = \SI{7450}{Hz}}$, ${R_{\mathrm{r}}+R_{\mathrm{mag}} = \SI{388.85}{k}\Omega}$, ${Q = 1121.7}$, ${S_{\mathrm{mag}} = \SI{80}{Hz/mT}}$, ${V_{\mathrm{mag}} = 6 \cdot 10^{-12}~\mathrm{m}^{3}}$) at room temperature (${T_0 = \SI{290}{K}}$). The fundamental LOD is limited by additive thermal-mechanical noise, thus $\mathrm{LOD}_{\mathrm{RrRmag}}$ (Eq.~\eqref{eq:LODRrRmagfx}) improves with the excitation amplitude $\hat{V}_{\mathrm{ex}}$. Parametric magnetically induced phase noise $S_{\varphi,\mathrm{M}}$ (Eq.~\eqref{eq:SvarphifxM2}) limits the LOD of real sensors. However, $\mathrm{LOD}_{\mathrm{M}}$ (Eq.~\eqref{eq:LODMfx1} and \eqref{eq:LODMfx2}) can be improved by decreasing the relative magnetic loss factor ${\tan \delta_{\mathrm{mag}}/\mu_{\mathrm{r,eff}}'}$.}
	\label{fig:lods_theory}
\end{figure}

\begin{figure}[t]
	\centering
	\includegraphics[width=0.5\textwidth]{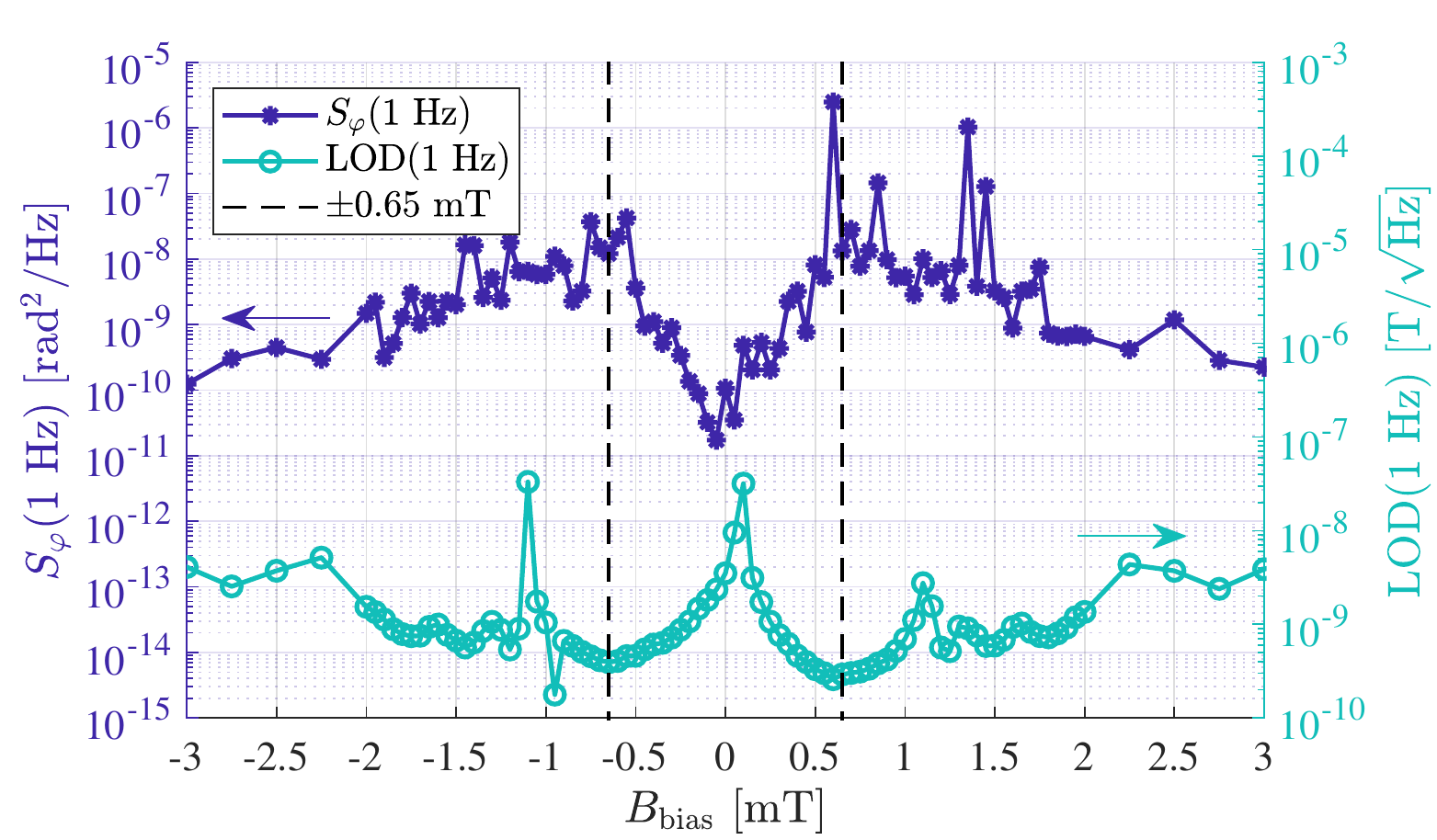}
	\caption{Measured phase noise at a frequency of \SI{1}{Hz} (same data as in Fig.~\ref{fig:phase_noise_bbias_sweep_bbias}) and measured limit of detection (LOD), also at a frequency of \SI{1}{Hz} for a constant excitation amplitude of ${\hat{V}_{\mathrm{ex}} = \SI{100}{mV}}$ as a function of $B_{\mathrm{bias}}$ measured increasingly from negative to positive magnetic saturation. Best values as low as ${\mathrm{LOD}(\SI{1}{Hz}) = 292~\mathrm{pT}/\sqrt{\mathrm{Hz}}}$ are obtained in the sensor's magnetic operating point at ${B_{\mathrm{bias}} = \SI{0.65}{mT}}$.}
	\label{fig:phase_noise_lod_bbias_sweep}
\end{figure}

The limit of detection (LOD) of a magnetic field sensor system denotes the frequency-dependent noise floor, i.e. an amplitude spectral density, in units of ${\mathrm{T}/\sqrt{\mathrm{Hz}}}$ \cite{Dur19a}. Thus, the LOD is given by the ratio of the amplitude spectral density of random phase fluctuations and the phase sensitivity.

Considering only the phase noise $S_{\varphi,\mathrm{RrRmag}}$ (Eq.~\eqref{eq:SvarphiRrRmagfx}) due to the resonator's thermal-mechanical noise, the fundamental LOD is given by
\begin{align}
	\mathrm{LOD}_{\mathrm{RrRmag}}(f_{\mathrm{x}}) &= \frac{\sqrt{S_{\varphi,\mathrm{RrRmag}}(f_{\mathrm{x}})}}{|S_{\mathrm{PM}}(f_{\mathrm{x}})|}\\
	&= \frac{f_{\mathrm{res}} \sqrt{4 k_{\mathrm{B}} T_0 (R_{\mathrm{r}} + R_{\mathrm{mag}})}}{\hat{V}_{\mathrm{ex}} Q |S_{\mathrm{mag}}|}
	\label{eq:LODRrRmagfx}
\end{align}
which is frequency-independent, deteriorates with the losses, and improves with the magnetic sensitivity. In particular, this fundamental LOD could be improved simply by increasing the excitation amplitude because the phase noise due to thermal-mechanical noise is additive (Fig.~\ref{fig:phase_noise_sweep_at_bsat}). For the sensor under investigation, with typical values (compare Fig.~\ref{fig:admittances_magnitude_of_f}) of ${f_{\mathrm{res}} = \SI{7450}{Hz}}$, ${R_{\mathrm{r}}+R_{\mathrm{mag}} = \SI{388.85}{k}\Omega}$, ${Q = 1121.7}$, and ${S_{\mathrm{mag}} = \SI{80}{Hz/mT}}$ the frequency-independent LOD would result, e.g. in a value as low as ${6.5~\mathrm{pT}/\sqrt{\mathrm{Hz}}}$ for an electrical driving amplitude of ${\hat{V}_{\mathrm{ex}} = \SI{1}{V}}$ (dashed line in Fig.~\ref{fig:lods_theory}).

However, because the sensor system's overall noise is dominated by magnetically induced phase noise (as shown in Fig.~\ref{fig:phase_noise_bbias_sweep_bbias} this noise is usually distinctly higher than phase noise due to the resonator's thermal-mechanical noise when the magnetic material is not saturated) it is more convenient to consider $S_{\varphi,\mathrm{M}}(f_{\mathrm{x}})$ (Eq.~\eqref{eq:SvarphifxM2}) for the determination of the detectivity
\begin{align}
	\mathrm{LOD}_{\mathrm{M}}(f_{\mathrm{x}}) &= \frac{\sqrt{S_{\varphi,\mathrm{M}}(f_{\mathrm{x}})}}{S_{\mathrm{PM}}(f_{\mathrm{x}})}\\
	&= \sqrt{\frac{4 k_{\mathrm{B}} T_0}{2 \pi f_{\mathrm{x}}~V_{\mathrm{mag}}}~\frac{\mu_0 \mu_{\mathrm{r,eff}}''}{(\mu_{\mathrm{r,eff}}')^2}}
	\label{eq:LODMfx1}
\end{align}
which improves with ${1/\sqrt{f_{\mathrm{x}}}}$. Remarkably, in that case, the limit of detection does not depend on the sensor's sensitivity at all but is solely determined by the volume and the magnetic properties $\mu_{\mathrm{r,eff}}'$ and $\mu_{\mathrm{r,eff}}''$ of the magnetostrictive film which, in turn, depend on the sensor's operating point in terms of bias field, excitation frequency, and excitation power. A basically identical result was reported e.g. for giant magnetoimpedance sensors, for which the fundamental detectivity is also independent of the sensitivity \cite{Mel08}. A recently published article \cite{Urs20} about magnetic domain activities confirms the relation between magnetic losses and magnetic noise in periodically driven magnetoelectric cantilevers. The authors also come to the conclusion that controlling the magnetic domain behavior is the key to optimum sensor performance.

For the sensor under investigation magnetically coated with a volume of $V_{\mathrm{mag}} = \SI{3}{mm} \cdot \SI{1}{mm} \cdot 2~\mu\mathrm{m} = 6 \cdot 10^{-12}~\mathrm{m}^{3}$, the expression for the LOD can be further simplified to
\begin{align}
	\mathrm{LOD}_{\mathrm{M}}(f_{\mathrm{x}}) &= \frac{\SI{23.1}{nT}}{\sqrt{f_{\mathrm{x}}}} \frac{\sqrt{\mu_{\mathrm{r,eff}}''}}{\mu_{\mathrm{r,eff}}'}\\
	&= \frac{\SI{23.1}{nT}}{\sqrt{f_{\mathrm{x}}}} \sqrt{\frac{\tan \delta_{\mathrm{mag}}}{\mu_{\mathrm{r,eff}}'}},
	\label{eq:LODMfx2}
\end{align}
clarifying the exclusive dependence on the magnetostrictive film's magnetic properties. The best value for the detectivity of ${\mathrm{LOD}(\SI{1}{Hz}) = 292~\mathrm{pT}/\sqrt{\mathrm{Hz}}}$ is measured around a bias flux density of ${B_{\mathrm{bias}} = \SI{+0.65}{mT}}$ (Fig.~\ref{fig:phase_noise_lod_bbias_sweep}) despite the higher losses compared to the operating point around ${B_{\mathrm{bias}} = \SI{-0.65}{mT}}$ (Fig.~\ref{fig:phase_noise_bbias_sweep_bbias}) for which a value of ${\mathrm{LOD}(\SI{1}{Hz}) = 394~\mathrm{pT}/\sqrt{\mathrm{Hz}}}$ is achieved. The reason is the higher magnetic sensitivity in this measurement of ${S_{\mathrm{mag}} = \SI{61.3}{Hz/mT}}$ at ${B_{\mathrm{bias}} = \SI{+0.65}{mT}}$ compared to a value of ${S_{\mathrm{mag}} = \SI{42.5}{Hz/mT}}$ at ${B_{\mathrm{bias}} = \SI{-0.65}{mT}}$. Thus, an optimum LOD is achieved at an operating point at which the sensitivity-to-loss ratio ${S_{\mathrm{mag}}/R_{\mathrm{mag}}}$ is maximized.

From a measured value ${\mathrm{LOD}(\SI{1}{Hz}) = 292~\mathrm{pT}/\sqrt{\mathrm{Hz}}}$ at ${B_{\mathrm{bias}} = \SI{+0.65}{mT}}$ the relative magnetic loss factor can be determined to ${\tan \delta_{\mathrm{mag}}/\mu_{\mathrm{r,eff}}' = 1.6 \cdot 10^{-4}}$ (${\tan \delta_{\mathrm{mag}}/\mu_{\mathrm{r,eff}}' = 2.9 \cdot 10^{-4}}$ at ${B_{\mathrm{bias}} = \SI{-0.65}{mT}}$). Due to the dependence of the magnetic properties on e.g. the material composition, thickness, magnetic domain configuration, shape, etc. and also due to their interdependencies it is difficult to compare the determined value with other values from the literature. However, values reported in \cite{Dur93} are at least in the same order of magnitude even though the investigated samples were measured at cryogenic temperatures. For typical parameters of the sensor under investigation, Fig.~\ref{fig:lods_theory} depicts resulting limits of detection at a frequency of \SI{1}{Hz} for various relative magnetic loss factors. Because the LOD is proportional to the square root of this loss factor, ${\tan \delta_{\mathrm{mag}}/\mu_{\mathrm{r,eff}}'}$ needs to be decreased by two orders of magnitude in order to improve the LOD by a factor of ten.

\section{Conclusion}
\label{sec:conclusion}
In this paper, a cantilever-type magnetoelastic resonant sensor, representative for other kinds of magnetoelastic resonators, has been investigated. Such sensors for the detection of low-frequency and low-amplitude magnetic fields utilize the $\Delta$E effect which leads to a magnetically induced resonance detuning. For the detection of the resonator's detuning, the sensor is preferably driven by an electrical excitation signal which, in turn, is then phase modulated by the magnetic measurement signal. Based on the dynamics of resonant mechanical structures an expression for the overall phase sensitivity has been derived. Such sensors exhibit several loss mechanisms that lead to random vibrations of the structure (thermal-mechanical noise) as well as to random agitation of the charge carriers flowing through the sensor (thermal-electrical noise). The phase noise resulting from these thermal noise sources can not only be predicted accurately but also decreased easily by increasing the excitation amplitude (additive noise). However, it has been shown that losses appearing in the sensor's magnetic material due to domain wall actions clearly generate additional flicker phase noise that can not be decreased by increasing the excitation amplitude (parametric noise). Based on the fluctuation-dissipation theorem indicating random fluctuations of the magnetization, an analytical expression for the magnetically induced phase noise could be derived. With this result, not only the fundamental LOD due to thermal vibrations of the mechanical structure but also the LOD for sensors impaired by magnetically induced phase noise could be described. In particular, in the latter case, the LOD does not depend on the sensitivity but is solely determined by the dynamic loss properties of the magnetic layer, at least if the magnetic sensitivity is high enough such that thermal noise sources are negligible. Hence, instead of the sensitivity, the magnetic losses, represented by the material's effective complex permeability, should be considered as the most important parameter for the further improvement of such sensors. This implication is not only valid for magnetoelastic cantilevers but also applies to any type of magnetoelastic resonator.

\begin{acknowledgments}
This work was supported (1) by the German Research Foundation (Deutsche Forschungsgemeinschaft, DFG) through the Collaborative Research Centre CRC 1261 \textit{Magnetoelectric Sensors: From Composite Materials to Biomagnetic Diagnostics}, (2) by the ANR Programme d'Investissement d'Avenir (PIA) under the Oscillator IMP project and the FIRST-TF network, and (3) by grants from the R\'{e}gion Bourgogne Franche-Comt\'{e} intended to support the PIA.
\end{acknowledgments}

\bibliography{mybibfile}

\end{document}